\documentclass[11pt]{article}
\pdfoutput=1
\usepackage{amsfonts}       
\usepackage{amsmath}        
\usepackage{amssymb}
\usepackage{epsfig}
\usepackage{amsthm}
\usepackage{amscd}
\usepackage{amstext}
\def\ben{\begin{equation}}
\def\een{\end{equation}}

\let\vp=\varphi

\let\w=\omega

\let\pa=\partial
\def\be{\begin{equation}}
\def\ee{\end{equation}}
\def\ba{\begin{array}}
\def\ea{\end{array}}

\def\vp{\varphi}

\def\dalemb#1#2{{\vbox{\hrule height .#2pt
        \hbox{\vrule width.#2pt height#1pt \kern#1pt
                \vrule width.#2pt}
        \hrule height.#2pt}}}

\newcommand{\bea}{\begin{eqnarray}}
\newcommand{\eea}{\end{eqnarray}}

\def\R{{{\Bbb R}}}

\def\Lag{{\mathcal{L}}}

\def\Op{{\mathcal{O}}}
\def\ocal{{\mathcal{O}}}

\def\calD{{\mathcal{D}}}

\def\charge{{q}}
\def\edensity{{\epsilon}}

\begin{document}

\begin{flushright}
NSF-KITP-08-123\\
PUPT-2278\\
arXiv:0810.1563 [hep-th]
\end{flushright}

\begin{center}
\vspace{1cm} { \LARGE {\bf Holographic Superconductors}}

\vspace{1.1cm}

Sean A. Hartnoll$^\flat$, Christopher P. Herzog$^\sharp$ and Gary
T. Horowitz$^\natural$

\vspace{0.7cm}

{\it $^\flat$ Jefferson Physical Laboratory, Harvard University\\
     Cambridge, MA 02138, USA }

\vspace{0.7cm}

{\it $^\sharp$ Department of Physics, Princeton University \\
     Princeton, NJ 08544, USA }

\vspace{0.7cm}

{\it $^\natural$ Department of Physics, UCSB \\
     Santa Barbara, CA 93106, USA }

\vspace{0.7cm}

{\tt hartnoll@physics.harvard.edu, cpherzog@princeton.edu, gary@physics.ucsb.edu} \\

\vspace{1.5cm}

\end{center}

\begin{abstract}
\noindent
It has been shown that a gravitational dual to a superconductor
can be obtained by coupling anti-de Sitter gravity to a Maxwell
field and charged scalar. We review our earlier
analysis of this theory and extend it in two directions. First, we consider all values
for the charge of the scalar field. Away from the large charge
limit, backreaction on the spacetime metric is important.
While the qualitative behaviour of the dual superconductor is found to be similar for
all charges, in the limit of arbitrarily small charge a new type of black hole instability
is found. We
go on to add a perpendicular magnetic field $B$ and obtain the London equation
and magnetic penetration depth. We show that these holographic
superconductors are Type II, i.e., starting in a normal phase at large $B$ and low
temperatures, they develop superconducting droplets as $B$ is
reduced.
\end{abstract}

\pagebreak

\tableofcontents

\setcounter{page}{1}
\setcounter{equation}{0}

\pagebreak

\section{Introduction}

Holographic superconductors are strongly coupled field theories which undergo
a superconducting phase transition below a critical temperature, and which
have a gravity dual in the sense of the AdS/CFT correspondence \cite{Maldacena:1997re}.
The reason for considering these systems is that they admit a large $N$ limit
in which many aspects of the physics can be studied directly, despite the theory being
inherently strongly coupled.\footnote{%
 The large $N$ limit in the best understood cases of AdS/CFT is
 the large $N$ limit of a gauge theory (with order $N^2$ ultraviolet degrees of freedom) rather than a  
 vector model (with order $N$ degrees of freedom). Unlike vector models,
 these theories do not become Gaussian in the large $N$ limit.
}
The AdS/CFT correspondence applied to these systems
therefore provides a tractable model for non-standard dynamical mechanisms
driving superconductivity.

The existence of holographic superconductors was established in
\cite{Gubser:2008px, Hartnoll:2008vx}. From the ($d$ dimensional) field
theory point of view, superconductivity is characterised by the condensation
of a, generically composite, charged operator $\ocal$ for low temperatures $T < T_c$.
In the dual ($d+1$ dimensional) gravitational description of the system, the transition to
superconductivity is observed as a classical instability of a black hole in
anti-de Sitter (AdS) space against perturbations by a charged scalar field $\psi$.
The instability appears when the black hole has Hawking temperature $T=T_c$.
For lower temperatures the gravitational dual is a black hole with a nonvanishing profile
for the scalar field $\psi$. The AdS/CFT correspondence relates the highly
quantum dynamics of the `boundary' operator $\ocal$ to simple classical dynamics of the
`bulk' scalar field $\psi$ \cite{Gubser:1998bc, Witten:1998qj}.

In this paper we will expand on previous works by considering two
aspects of holographic superconductors in $d=3$ spacetime
dimensions in some depth. Firstly, our previous computations of
the conductivities in the superconducting phase
\cite{Hartnoll:2008vx} were performed in a limit in which the
charge of the operator $\ocal$ was taken to be large. This limit
had the virtue of eliminating the backreation of the scalar field
$\psi$ on the spacetime metric. We had expected this limit to
 capture accurately the essential physics and, by considering the
theory with backreaction onto the metric,
 we show below that indeed it does. We will furthermore discuss the full
behaviour of the theory as a function of the charge of $\ocal$.
Remarkably, we find that superconductivity persists for
arbitrarily small charge. Even when $\ocal$ is neutral, a
condensate forms at low temperature. This last fact indicates that
there are two different mechanisms driving the instability, as we
will discuss below.

Backreation on the metric will result in a coupling between the electric and energy
currents. In addition to electrical conductivity, we also compute thermal and thermoelectric conductivities.
Given that our computations are in the clean limit --- there are no impurities
\cite{Hartnoll:2008hs} --- our system is translationally invariant. This results in the
Drude peak becoming a divergence in the electrical conductivity at
$\w=0$, even in the normal phase. The divergence due to
translational invariance (i.e. conservation of momentum) should be
distinguished from the symmetry breaking infinite
superconductivity current, as we discuss in some detail below.

In the second half of the paper we shall study the behaviour of
the superconducting phase under an external magnetic
field.\footnote{%
 There have been a couple of previous discussions of holographic
 superconductors in the presence of magnetic fields
 \cite{Albash:2008eh, Maeda:2008ir}. These
 have not settled the question of the type of superconductor. Some early
 attempts to add magnetic fields are \cite{Nakano:2008xc, Wen:2008pb}.
 } 
The theories that we are considering do not have a
dynamical photon; the $U(1)$ symmetry that is spontaneously broken
in the superconducting phase is a global symmetry. At zeroth order
the theory is perhaps more accurately described as a charged
superfluid. One might therefore have suspected that little could
be said about the crucial magnetic properties of superconductors,
such as the Meissner effect and the difference between type I and
type II superconductors. However, we shall show that pessimism is
unjustified. The key aspect of the Meissner effect involves the
generation of currents as described by the London equation. We see
that this generation indeed occurs in holographic superconductors.
Furthermore, by weakly gauging the theory we explicitly determine
the type of the superconductor and compute the mass generated for
the photon.

The symmetry breaking condensates that we study in this paper all
have an $s$ wave character. It is known that the AdS/CFT
correspondence can also describe condensates that appear to share
many properties of $p$ wave superconductors \cite{Gubser:2008zu,
Gubser:2008wv, Roberts:2008ns}. It would be interesting to address
the range of questions we consider below for the $p$ wave duals.

Since we are using AdS/CFT to describe superconductivity, it is natural to ask what is the connection between superconductivity and conformal field theories. One connection is that the CFT describes a quantum critical point, i.e., a phase transition at zero temperature. In some cases the large fluctuations associated with the quantum phase transition can induce the pairing responsible for superconductivity \cite{belitz}. The quantum critical theory is the theory of these fluctuations, and AdS/CFT will provide us with models for such theories. In this work however, we simply use the fact that the AdS/CFT correspondence has been extended to nonconformal field theories as well. We will break the conformal invariance by adding a background charge density.

This paper is organized as follows. In the next section we write down the equations for the gravity dual of a superconductor. The rest of the paper is devoted to investigating  the solutions to these equations. In section three we study the static solutions describing hairy black holes which are dual to the superconducting phase with nonzero condensate. The following section contains a discussion of transport phenomena in the superconductor, using perturbations of the black hole. In section five, we introduce an orthogonal magnetic field and show that the superconductor is Type II. The next section contains a discussion of the currents induced by the magnetic field, and shows how the London equation is recovered. In section seven we explicitly weakly gauge the theory and compute the
photon mass in the superconducting phase. Some comments on the relation and difference between the holographic approach to superconductivity and Landau-Ginzburg theory are in section eight. We conclude with a short summary and some open questions.

\setcounter{equation}{0}
\section{The bulk equations for a holographic superconductor}
\label{sec:bulk}

The best understood examples of the AdS/CFT correspondence involve
AdS spaces that are part of a full ten or eleven dimensional
solution to string or M theory. The low energy fields that
propagate in the AdS space are obtained as consistent truncations
of the higher dimensional theory. However, in the field of
`applied AdS/CFT' (one is thinking usually of applications to QCD
or condensed matter physics) a different philosophy is
possible. A phenomenological approach is taken in which the
classical fields which propagate in the bulk and their
interactions are chosen by hand to capture the physics
of interest. The two main drawbacks of this approach are that
firstly it is not possible to compute many quantities away from
the large $N$ limit (because the bulk theory is likely not part of
a consistent theory of quantum gravity) and secondly we do not
have an explicit description of the field content and interactions
of the dual field theory. We believe it will be possible and
certainly interesting to realise holographic superconductors as
truncations of string theory. (A concrete suggestion for embedding
$p$ wave superconductors was made in \cite{Roberts:2008ns}.) For
the moment we will work with a phenomenological model.

What are the minimal ingredients we need to describe a holographic
superconductor? We are interested in the continuum limit, as
AdS/CFT has not yet been developed for lattices, and therefore our
field theory will have a conserved energy momentum tensor
$T^{\mu\nu}$. For simplicity we will be working with a theory that
is Lorentz invariant at high energies and so the indices $\mu,\nu$
run over $t,x,y$. The AdS/CFT correspondence has recently been
generalised to non-relativistic theories \cite{Son:2008ye,
Balasubramanian:2008dm}, and it would certainly be interesting to
adapt our study to those cases. Furthermore we need a global
$U(1)$ symmetry in the field theory, and therefore we will have a
conserved current $J^\mu$. Finally, as we wish to break this
$U(1)$ symmetry spontaneously, we need a charged operator $\ocal$,
which will condense at low temperature.

The most basic entries in the AdS/CFT dictionary \cite{Gubser:1998bc, Witten:1998qj} tell us that
there is a mapping between field theory operators and fields in the bulk. In particular, $T^{\mu\nu}$ will
be dual to the bulk metric $g_{ab}$, the current $J^\mu$ will be dual to a Maxwell field in the bulk, $A_a$, whereas the operator $\ocal$ will be dual to a charged scalar field $\psi$ (which is therefore necessarily complex). Here $a,b$ run over the four bulk coordinates $t,x,y,r$. The next step is to write down a minimal Lagrangian involving these fields. The Lagrangian density for a Maxwell field and a charged complex scalar field coupled to gravity is
\be\label{eq:bulktheory}
\Lag = R + \frac{6}{L^2} - \frac{1}{4} F^{ab} F_{ab} - V(|\psi|)
- |\nabla \psi - i \charge A \psi |^2 \,.
\ee
As usual we are writing $F=dA$.
Most of our work will revolve around solving the equations of motion
that are obtained from this Lagrangian. These are the scalar equation
\be
- \left(\nabla_a - i \charge A_a \right) \left(\nabla^a - i \charge A^a
\right)\psi + \frac{1}{2} \frac{\psi}{|\psi|} V'(|\psi|) = 0 \,,
\ee
Maxwell's equations
\be
\nabla^a F_{ab} = i \charge \left[\psi^* (\nabla_b - i \charge A_b)\psi - \psi (\nabla_b + i \charge A_b)\psi^* \right]
\,,
\ee
and Einstein's equations
\bea\label{einsteineq}
\lefteqn{R_{ab} - \frac{g_{a b} R}{2} - \frac{3 g_{a b}}{L^2} =
\frac{1}{2} F_{a c} F_b{}^c - \frac{g_{a b}}{8} F^{cd} F_{cd} - \frac{g_{a b}}{2} V(|\psi|)}
\nonumber \\
& \displaystyle{  - \frac{g_{a b}}{2}
|\nabla
\psi - i \charge A \psi|^2 + \frac{1}{2} \left[(\nabla_a \psi - i \charge A_a
\psi) (\nabla_b
\psi^* + i \charge A_b \psi^*) + a \leftrightarrow b \right]  \,.}
\eea
Note in these equations that $\charge$ is the charge of the scalar
field.

The first step will  be to find static black hole solutions to
these equations. These solutions describe the equilibrium phases
of the theory. It is in these solutions that we see a phase
transition as the temperature is lowered. Once we have the
equilibrium solutions we can look at perturbations away from
equilibrium, which describe transport processes.

\setcounter{equation}{0}
\section{Normal and superconducting phases}
\label{sec:phases}

\subsection{The hairy black hole ansatz}

In the AdS/CFT correspondence, plasma or fluid-like phases of the field theory at nonzero temperature are described by black hole solutions to the bulk gravitational action \cite{Witten:1998zw}. 
To study conductivity and other transport properties of a charged plasma, 
the first step therefore is to find black hole solutions to our theory (\ref{eq:bulktheory}). 
Superfluidity and superconductivity are associated with symmetry breaking, and thus we search for
solutions in which the charged scalar field has a nontrivial expectation value.

One further ingredient is necessary. The simplest theories described by the AdS/CFT correspondence are conformally (in particular, scale) invariant. In a conformal field theory in Minkowski space, such as ours, in the absence of another scale, all nonzero temperatures are equivalent.
Therefore, if we wish to obtain phase transitions at a critical temperature,
we need to introduce another scale.
Among the various scales to choose from,
we introduce a finite charge density $\rho$ (equivalently, a finite chemical potential $\mu$). In 2+1 spacetime dimensions, $\rho$ has dimensions of mass squared whereas $\mu$ has mass dimension one.
Our primary motivation for introducing a scale in this manner is the observation in \cite{Gubser:2008px} that it leads to a superconducting instability at low temperatures. If we were to compare with real experimental systems, our model describes a quantum critical theory that has been deformed by doping \cite{Hartnoll:2007ih}. Another possible connection is to relativistic systems such as graphene held at finite gate voltage.

As we shall review below, a charge density $\rho$ in the system corresponds to giving an electric charge to the black hole. We shall consider magnetic charges also in a later section.
The upshot is that we are looking for electrically charged plane-symmetric hairy black hole solutions.
Thus we take the metric ansatz
\be
\label{eq:metric}
ds^2 = - g(r) e^{-\chi(r)} dt^2 + \frac{dr^2}{g(r)} + r^2 \left(dx^2 + dy^2
\right)
\,,
\ee
together with
\be
\label{eq:Aandpsi}
A = \phi(r) dt \,, \qquad \psi = \psi(r) \,.
\ee
We now look for solutions to the above equations of motion with this form.

It is immediately seen that the $r$ component of Maxwell's
equations implies that the phase of $\psi$ must be constant.
Without loss of generality we therefore take $\psi$ to be real for
the background. The scalar equation becomes
\be\label{scalareq}
\psi'' + \left(\frac{g'}{g} -{\chi'\over 2} + \frac{2}{r} \right) \psi' + \frac{\charge^2 \phi^2 e^\chi}{g^2} \psi -
\frac{1}{2g} V'(\psi) = 0 \,,
\ee
Maxwell's equations become
\be\label{eq:max}
\phi'' + \left({\chi'\over 2} + \frac{2}{r} \right)
\phi' - \frac{2 \charge^2 \psi^2}{g} \phi = 0 \,,
\ee
while the $tt$ and $rr$ components of Einstein's equations yield
\bea
\chi' + r\psi'^2  + {r \charge^2 \phi^2\psi^2e^\chi\over g^2 }=0 \label{eq:Eone}\,, \\
\frac{1}{2} \psi'^2 +  \frac{\phi'^2e^\chi}{4g} + \frac{g'}{g
r}+ \frac{1}{r^2} - \frac{3}{g L^2} +
\frac{V(\psi)}{2 g} + \frac{\charge^2 \psi^2 \phi^2 e^\chi}{2 g^2} = 0 \,.
\label{eq:Efinal}
\eea
Note that both these equations only have first derivatives,
although they appear squared.  The $xx$ component of Einstein's
equations is not independent and follows from differentiating the
two equations above. We will specialise for concreteness to the
simple potential
\be\label{eq:mass}
V(\psi) = - \frac{2}{L^2} \psi^2 \,.
\ee
This is the conformal mass term for  a scalar in $AdS_4$ and is above the
Breitenlohner-Freedman bound for stability. We choose this value following
\cite{Hartnoll:2008vx}. The effect of other masses is discussed in \cite{Horowitz}.
A stringy embedding of this model would of course fix the full potential.

If one takes the limit $\charge \rightarrow \infty$ keeping
$\charge \psi$ and $\charge \phi$ fixed, the matter sources drop
out of Einstein's equations (\ref{eq:Eone}) and (\ref{eq:Efinal}),
while the scalar and Maxwell equations
(\ref{scalareq},\ref{eq:max}) remain essentially unchanged. This
is the probe limit studied in
\cite{Hartnoll:2008vx}. Our first objective in this paper is to go beyond
the probe limit. We will solve the full set of equations (with finite
$\charge$) numerically by integrating out from the horizon to
infinity. By considering a series solution at the horizon --- the
horizon radius $r_+$ is defined through the requirement that
$g(r_+) = 0$
--- one finds that there are four independent parameters at the
horizon
\be
r_+ \,, \quad \psi_+ \equiv \psi(r_+) \,, \quad E_+ \equiv
\phi'(r_+) \,, \quad \chi_+ = \chi(r_+) \,.
\ee
The third of these quantities is the value of the electric field
at the horizon. The scalar potential $\phi$ itself must go to zero
at the horizon in order for the gauge connection to be regular.
These quantities determine the Hawking temperature of the black
hole (for instance, from regularity of the Euclidean solution)
\be
T = \left((12 + 4 \psi_+^2) e^{-\chi_+/2} - L^2 E_+^2 e^{\chi_+/2}
\right) \frac{r_+}{16 \pi L^2} \,.
\ee

At infinity we have the following parameters that determine the
charges of the black hole and the expectation values of scalar
fields. The charge density, $\rho$, and chemical potential, $\mu$,
are read off \cite{Gubser:1998bc, Witten:1998qj}
(more explicitly, see for instance \cite{Hartnoll:2007ai})
from the asymptotic value of the scalar potential as
$r \to \infty$
\be\label{eq:phiasymptotic}
\phi = \mu - \frac{\rho}{r} + \cdots \,.
\ee
The general asymptotic behaviour of the scalar field as $r \to
\infty$ is
\be \label{eq:psiasymptotic}
\psi = \frac{\psi^{(1)}}{r} + \frac{\psi^{(2)}}{r^2} + \cdots  \,.
\ee
This simple falloff is another reason for choosing the mass
(\ref{eq:mass}). This value of the mass falls within the range in which
there is a choice of admissible boundary conditions at large radius
\cite{Klebanov:1999tb}. Depending  on the choice of boundary conditions,
we can read off the expectation value of an operator $\ocal_2$, of
mass dimension two, or of an operator $\ocal_1$, of mass dimension
one. Specifically, for a stable theory we must either impose
\be\label{eq:bc1}
\psi^{(1)} = 0 \,, \quad \text{and} \quad \langle \Op_2 \rangle
= \sqrt{2} \psi^{(2)} \,,
\ee
or
\be\label{eq:bc2}
\psi^{(2)} = 0 \,, \quad \text{and} \quad \langle \Op_1 \rangle
= \sqrt{2} \psi^{(1)} \,.
\ee
The factor of $\sqrt{2}$ is following \cite{Hartnoll:2008vx} and is a convenient
normalisation.

In order for the Hawking temperature of the black hole to be the
temperature of the boundary field theory, we must impose
\be
\chi \to 0 \,, \quad \text{as} \quad r \to \infty \,.
\ee
This is a statement about the normalisation of the time coordinate $t$
relative to the gravitational redshift, as determined by the normalisation of the
$r$ coordinate.
In practice we can implement this boundary condition by taking an
arbitrary $\chi_+$ at the horizon, obtaining the asymptotic value of $\chi$,
and then rescaling time $t \to a t$, in order to set the
asymptotic value to zero.  Said another way, we are using the scaling symmetry of the metric, gauge field, and equations of motion
\be
e^{\chi} \to a^2 e^{\chi} \, , \quad t \to a t \, , \quad \phi \to \phi / a \ ,
\ee
to set $\chi=0$ at the boundary.

There are furthermore two scaling symmetries of the equations of motion
that we can use to
set $L=1$ and $r_+=1$ when performing numerics. The first is
\be
r \to a r \,, \quad t \to a t \,, \quad L \to a L \,, \quad \charge \to
\charge/a \,,
\ee
which rescales the metric by $a^2$ and $A = \phi \, dt$ by $a$. The second is
\be
r \to a r \,, \quad (t,x,y) \to (t,x,y)/a \,, \quad g \to a^2 g \,, \quad \phi \to a \phi
\,,
\ee
which leaves the metric and $A$ unchanged. 
After these scaling actions we are left with one parameter in the
Lagrangian which is physical, the charge of the scalar field $\charge$,
and two parameters that determine the initial data at the horizon,
$\psi_+$ and $E_+$. Integrating out from the horizon to infinity
gives a map
\be\label{eq:map}
(\psi_+, E_+) \mapsto (\mu, \rho, \psi^{(1)}, \psi^{(2)}, \edensity) \,.
\ee
We have included the mass of the black hole solution on the right
hand side of this expression. We have denoted the mass by
$\edensity$, as it is to be interpreted as the energy density of
the field theory. The mass is to be read off from the large $r$
behaviour of $g$ and $\chi$. Assuming that $\chi \to 0$ as $r\to
\infty$ we have
\bea
g &=& \frac{r^2}{L^2} + \frac{(\psi^{(1)})^2}{2 L^2} +
\frac{-\edensity L^2/2 + 4 \psi^{(1)}\psi^{(2)} /3L^2}{r} + \cdots \,,
\label{eq:gbryexp}
\\
e^{-\chi} g &=& \frac{r^2}{L^2} - \frac{\edensity L^2}{2r} + \cdots \,.
\label{eq:chibryexp}
\eea

Finally, upon imposing either of the boundary conditions
(\ref{eq:bc1}) or (\ref{eq:bc2}), the map (\ref{eq:map}) reduces
to a one parameter family of solutions for each value of the
scalar field charge $\charge$. We can think of this parameter as
being the temperature of the theory at a fixed charge density.
Depending on whether the scalar $\psi$ is nonzero or not for this
solution, we will be in the superconducting or normal phases,
respectively.

\subsection{Phase diagram}

\begin{figure}[h]
\centerline{a) \epsfig{figure=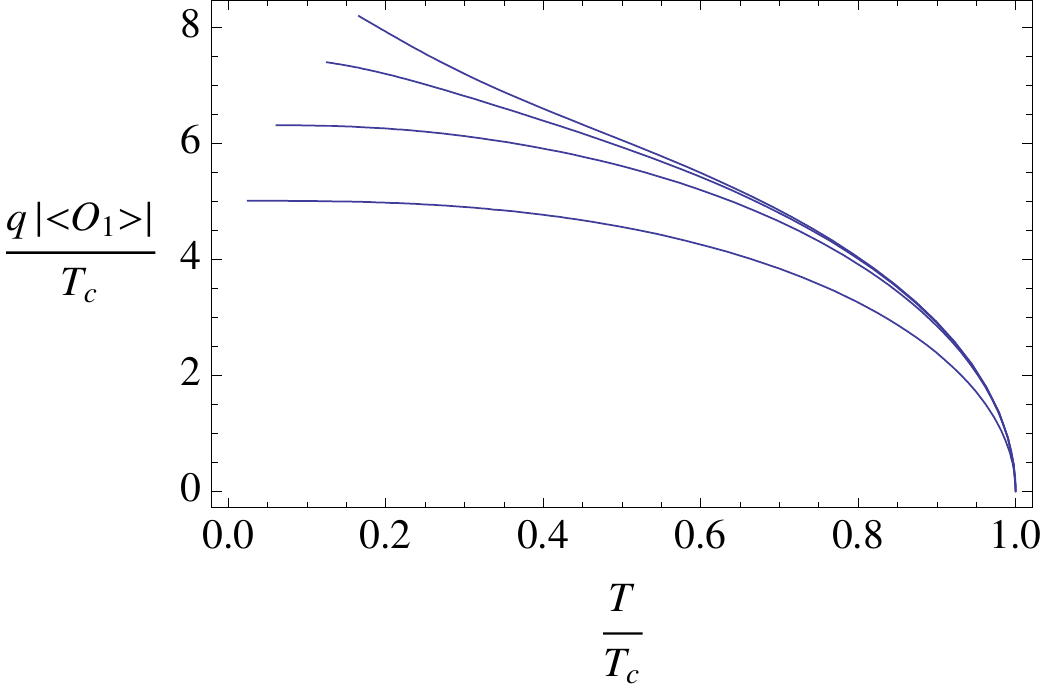, width=3in}  b) \epsfig{figure=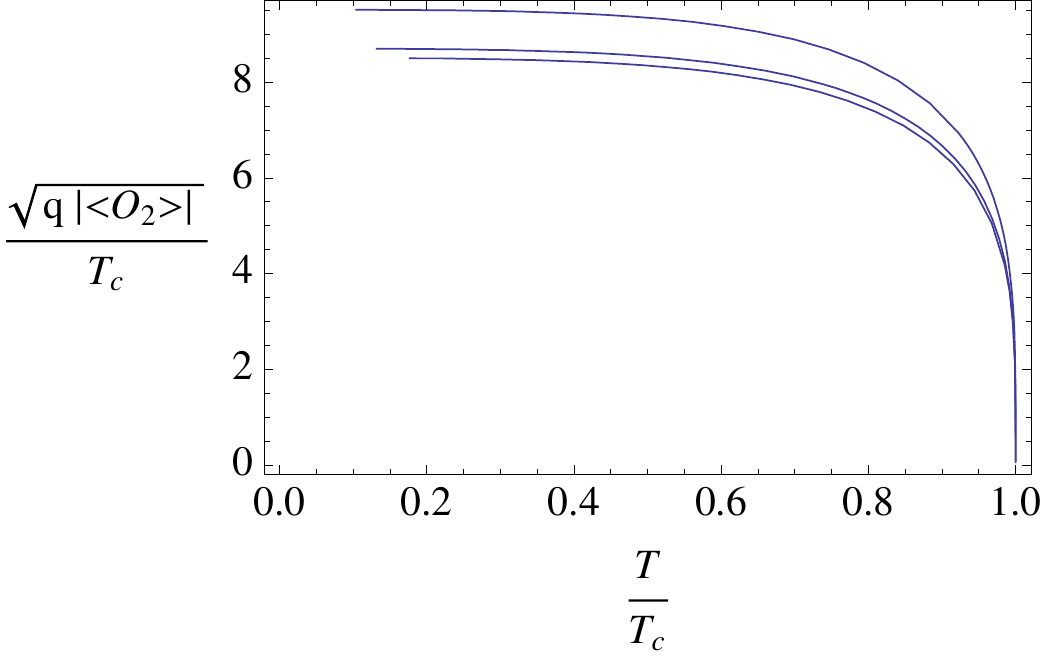, width=3in}}
\caption{
\label{condensate}
The value of the condensate as a function of temperature, with the charge density held fixed, for the
two different boundary conditions: a) from bottom to top, the various curves correspond to $\charge=1$, 3, 6, and 12; b) from top to bottom, the curves correspond to $\charge = 3$, 6, and 12. The value $q=1$
gives a much larger condensate in this case, achieving $\sqrt{q \langle \ocal_2 \rangle}/T_c \approx 21$ so we have not plotted it. Note that the large $q$ limit is approached in opposite directions in the two cases.}
\end{figure}

In figure \ref{condensate} we show the condensate for the charged scalar field as a function of temperature with the charge density held fixed. We plot $q\langle\ocal\rangle$ since this is the quantity which is finite in the probe (large $q$) limit. 
This plot has been obtained by numerically solving the differential equations in the previous
subsection using a shooting method. Although there is an ambiguity at this point in the
normalisation of the scalar field, we will shortly
relate the condensate, in our normalisation, to physical quantities such as the gap in the frequency dependent conductivity.

The most important feature of the plots is that in all cases there
is a critical temperature $T_c$ below which a charged condensate
forms. This is the symmetry breaking phase transition to a
superconducting phase. For $T > T_c$ the solution is simply the
Reissner-Nordstrom-AdS black hole. That is (with $L = 1$)
\be
\chi = \psi = 0 \,, \qquad g = r^2 - \frac{1}{r} \left(r_+^3 + \frac{\rho^2}{4 r_+} \right)
+ \frac{\rho^2}{4 r^2}  \,, \qquad \phi =  \rho
\left(\frac{1}{r_+} - \frac{1}{r} \right)\,.
\ee
At these high temperatures, there are no hairy back hole
solutions. At $T=T_c$ the Reissner-Nordstrom-AdS solution becomes
unstable against perturbations of the scalar field. As pointed out
in \cite{Gubser:2008px}, this instability can be understood directly from the
fact that the coupling of the scalar to the gauge field through
covariant derivatives induces an effective negative mass term for
the scalar field. This term becomes more important as the
temperature is lowered at fixed charge density, eventually driving
the scalar field tachyonic. For $T < T_c$ we find that hairy black
hole solutions do exist and have a lower free energy than the
`bald' black hole.

 The general form of the
curves is similar to the well known case of, for instance, BCS
theory. The condensate turns on at $T=T_c$ following a square root
law $\langle \ocal_i \rangle \sim T_c^i (1-T/T_c)^{1/2}$, as is
typical for mean field theory treatments of
second order transitions. As the temperature is taken
to zero, the condensate tends to a finite value in terms of the
scale set by $T_c$.  It is interesting to note the dependence of the condensate
on the charge of the operator $\ocal$. The most striking effect  is seen in
the theory with operator $\ocal_1$, in figure \ref{condensate}a. It had been
found in \cite{Hartnoll:2008vx} that in the probe limit ($q \to
\infty$) the condensate appeared to diverge as $T/T_c \to 0$, perhaps
indicating an instability of the theory. In figure \ref{condensate}a we can see
how the full backreacting system cures this divergence. At smaller
values of the charge $\charge$ there is no sign of a divergence.
As $\charge$ is increased the plot starts to curve upwards as we
approach low temperatures and appears similar to that of the probe
limit. Unfortunately, it is difficult to get the numerics reliably
down to very low temperatures so we have not been able to see the
resolution of the divergence at large but finite $\charge$
explicitly. Clearly it would be extremely desirable to have a more
direct approach to the zero temperature properties of holographic
superconductors. One numerical observation we can make is that
the electric field on the horizon, $E_+$, appears to go to zero
at low temperatures. Note that, in contrast,  for the $\ocal_2$ theory the condensate always stays bounded as $T/T_c \to 0$ and probe limit is approached from above. We believe that this behavior is more typical of holographic superconductors. 

As was noted in \cite{Hartnoll:2008vx} a second order transition is only possible in our 2+1
dimensional system at finite temperature
because we are working in a large $N$ limit. The large $N$ limit suppresses
fluctuations of the fields, in particular, the massless fluctuations associated to the spontaneous breaking of our global $U(1)$ symmetry. Away from the large $N$ limit, these fluctuations will lead to infrared
divergences which will destroy the long range order (the Coleman-Mermin-Wagner theorem) in the
low temperature phase. It would be extremely interesting to capture these fluctuations within an AdS/CFT framework and perhaps to exhibit an algebraic order with spatially separated correlations
falling off like $(\Delta x)^{-1/N^\#}$. A discussion of large $N$ expansions and symmetry breaking in two dimensions can be found, for instance, in \cite{Witten:1978qu}. A related question is whether any analogue of the Berezinski-Kosterlitz-Thouless transition can be seen in this or other AdS/CFT systems.
Finally, in section seven below we will gauge the $U(1)$ symmetry (with 2+1 dimensional photons)
and see that the Goldstone boson is eaten by the photon, which becomes massive.
Once there are no Goldstone bosons, there are no longer IR divergences.

The critical temperature $T_c$ appearing in figure \ref{condensate} is set by the
only other dimensionful scale in the system, the charge density
$\rho$. (In the grand canonical ensemble, 
we could alternatively consider the scale to be set by the
chemical potential $\mu$.) Dimensional analysis implies that we
will have $T_c \propto \sqrt{\rho}$. However, the constant of
proportionality will also depend on the charge $\charge$ of the
operator $\ocal$. This dependence is shown in figure 2.

\begin{figure}[h]
\centerline{a) \epsfig{figure=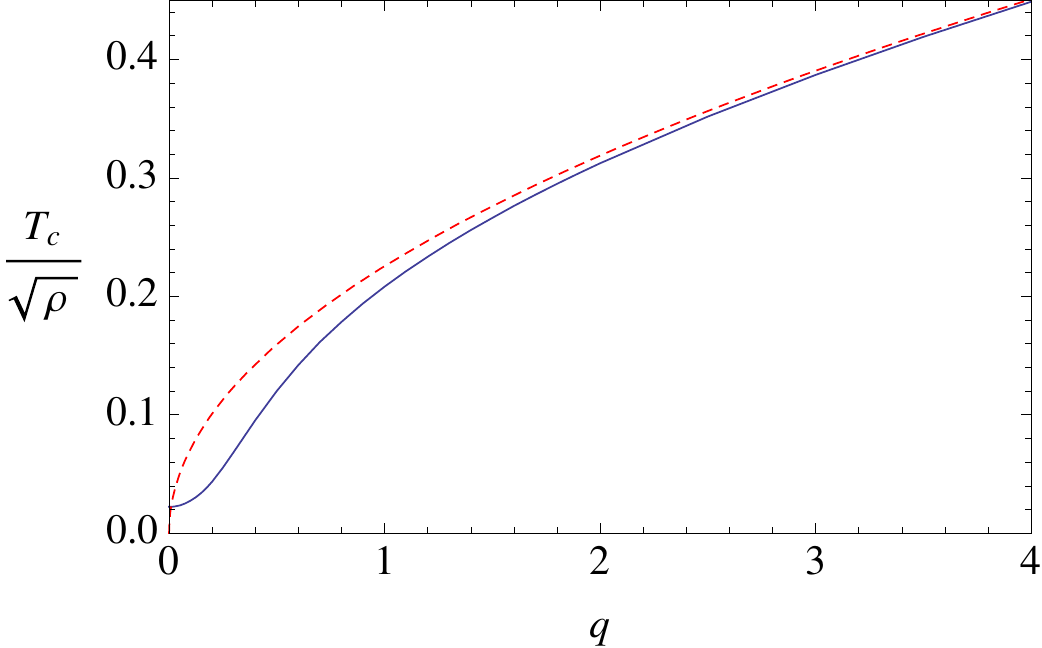, width=3in}  b) \epsfig{figure=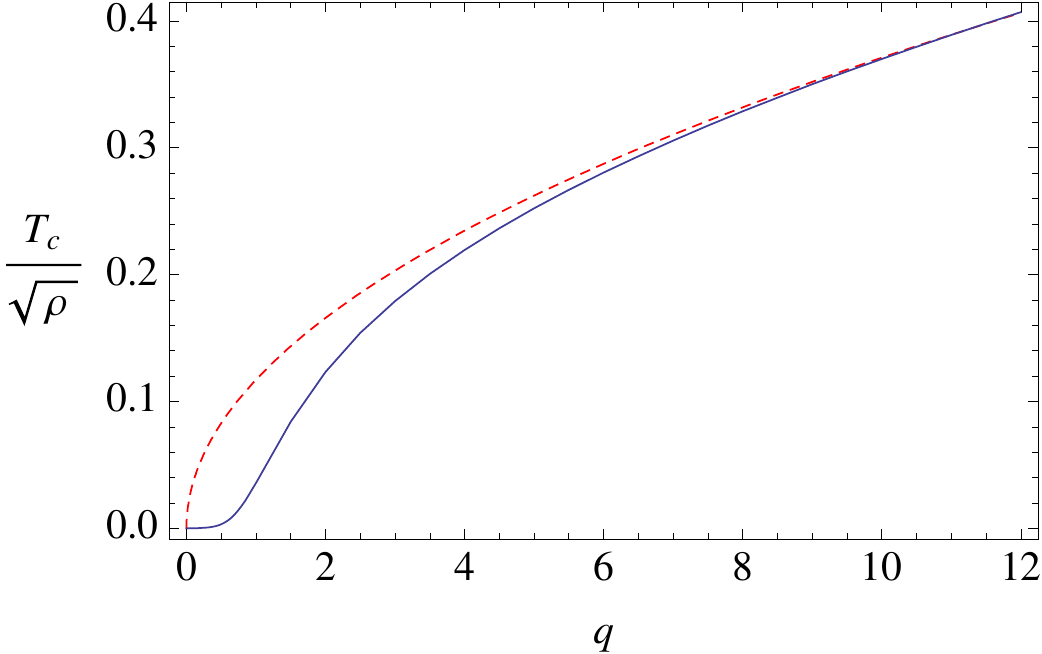, width=3in}}
\caption{
\label{Tc}
The blue solid line is the critical temperature as a function of
$\charge$.  The dashed red line is the probe limit (naively extrapolated to all $\charge$). a) The
dimension one case where the probe limit corresponds to $0.2255
\sqrt{\charge}$; b) the dimension two case with the probe limit
$0.1173 \sqrt{\charge}$.
 }
\end{figure}

In figure \ref{Tc} we have also included the value of $T_c$ in the probe (large $\charge$) limit. This was computed in
\cite{Hartnoll:2008vx} where it was found that $T_c \propto \sqrt{\charge \rho}$. The effects of backreaction
produce two changes with respect to the probe limit. Firstly,
$T_c$ is suppressed compared to the probe
estimate everywhere except at very small $\charge$.  
This suppression relative to the probe limit can be understood in part from the 
correction to the Hawking temperature of a black hole due to the electric 
charge: $T \sim 12 r_+^4 - \rho^2$.
Away from the probe limit, the electric charge back reacts on the metric, 
and this charge decreases slightly the temperature.
Secondly, $T_c$ remains nonzero for all $\charge$, including $q=0$.\footnote{%
 If $T_c$ were to go to zero
 at some finite $q$, we would have a zero temperature quantum phase transition at that point.
 We expect such a quantum critical point for more positive masses.
} 
In other words, even neutral
operators can condense at low temperature. Lest one doubt our
numerics, we give a proof of this fact in the appendix for the
case $\langle \ocal_1 \rangle \neq 0$.

\subsection{A new type of instability}

At first sight it is very surprising that a near extremal charged
black hole is unstable to forming neutral scalar hair. 
 As mentioned earlier,
the reason charged scalar hair is expected is that the coupling of
the scalar to the gauge field through covariant derivatives induces
an effective negative mass term for the scalar field. This term
becomes more important as the temperature is lowered at fixed
charge density, eventually driving the scalar field tachyonic.
This mechanism does not apply to neutral scalar fields, so the
origin of the hair in this case must be qualitatively different.

The best explanation seems to be the following.\footnote{%
 We thank
 M. Roberts for suggesting this.
} 
First we note that although
$T_c$ is not zero at $\charge = 0$, it is small. Therefore we will
think of these unstable black holes as being near extremal. An
extremal Reissner-Nordstrom AdS black hole has a near horizon
geometry which is $AdS_2\times \R^2$. Our neutral scalar has $m^2
= -2/L^2$. The Breitenlohner-Freedman (BF) bound governing stability
of scalar fields in $AdS_D$ is $m^2_{BF} = - (D-1)^2/4L^2$.  So
while our scalar field is above the BF bound for $AdS_4$, it is
below the BF bound for $AdS_2$. The $AdS_2$ radius of curvature is
actually smaller than $AdS_4$ ($L_2^2 = L_4^2/6$ for an extremal
black hole), but even taking this into account, our scalar is
below the BF bound in the near horizon region of the black hole.
This argument suggests that a Reissner-Nordstrom AdS black hole, when
coupled to a neutral scalar with $m^2 = -2/L^2$, becomes unstable near
extremality. A proof of this statement is given in Appendix \ref{app:unstable}.
The instability produces the hairy black holes we see
numerically.\footnote{%
 An earlier example of a charged AdS black hole with neutral scalar hair was given in  
 \cite{Martinez:2006an}. That example also involved a scalar with $m^2 = -2/L^2$;  
 however it required a black hole with a hyperbolic horizon.
}  
(This argument alone does not explain why the
dimension two case with faster fall off should be more stable,
which is what is seen numerically.) Note that extremal
Reissner-Nordstrom AdS black holes are not supersymmetric, so
their instability does not lead to a contradiction with general
stability results.

The physical point to take away from these observations is that
there are (at least) two distinct physical mechanisms leading to
superconductivity in our system. At very large charges for the
scalar fields, it is the (bulk) gauge covariant derivatives that
enhance the effective negative mass. At very small charges, it is
the fact that the near extremal charge of the black hole produces
a throat in which even a neutral scalar with sufficiently negative
mass squared becomes unstable. The crossover between these effects
presumably corresponds to the crossover visible in figure 2.
Although the crossover appears to be smooth, one might bear in mind
the possibility of a phase transition as the type of instability swaps.
It would be interesting, of course, to re-interpret this crossover
from the dual field theory.

\subsection{The hairy black hole action}

We conclude this section by  computing the Euclidean action for our hairy black hole: 
\be
S_E = - \int d^4x  \sqrt{-g_B} {\cal L}\,,
\ee
where ${\cal L}$ is given in (\ref{eq:bulktheory}) and $g_B$ is the determinant of the bulk metric. We first show that, when evaluated on a solution, this action reduces to a simple surface term at infinity. From the symmetries of the solution  (\ref{eq:metric}, \ref{eq:Aandpsi}), the $xx$ component of the stress energy tensor only has a contribution from the terms proportional to the metric. Thus, Einstein's equation (\ref{einsteineq}) implies that the Einstein tensor satisfies
\be
G_{xx} = {1\over 2} r^2({\cal L} - R) \,.
\ee
This implies
\be
-R = {G^a}_a = {G^t}_t + {G^r}_r + {\cal L} - R \,,
\ee
or 
\be
 {\cal L} = - {G^t}_t - {G^r}_r =  - {1\over r^2}\left [(rg)' + (rg e^{-\chi})' e^\chi\right] \,.
\ee
The Euclidean action is then a total derivative
\be
S_E  =\int d^3x  \int_{r_+}^{r_\infty} dr [2 rg e^{-\chi/2}]' \,.
\ee
The surface term on the horizon vanishes since $g(r_+)=0$. So we get just the surface term at ${r_\infty}$
\be
S_E =  \left. \int d^3x\  2rge^{-\chi/2} \right|_{r=r_\infty} \,.
\ee

This action diverges as ${r_\infty} \to \infty$ and must be regulated. The counter terms we need to regulate it are standard (see for example \cite{Skenderis:2002wp}).  We require a Gibbons-Hawking term and a boundary cosmological constant:
\be
S_\text{c.t.} =  \left. \int d^3 x \sqrt{-g_{\infty}} \left( - 2K + 4/L\right) \right|_{r=r_\infty} \ ,
\ee
where $g_{\infty}$ is the induced
metric on the boundary $r=r_\infty$ and $K = g_\infty^{\mu \nu} \nabla_\mu n_\nu$ is the trace of the extrinsic
curvature ($n^\mu$ is the outward pointing unit normal vector to the
boundary).
We also require a term quadratic in the scalar field
that depends on which boundary condition we choose for $\psi$.
%
If we fix the value of $\psi^{(1)}$ on the boundary, we must add
\be
S_1 = \left. \int d^3x \sqrt{-g_{\infty}}\,  \psi^2/L \right|_{r=r_\infty} \ ,
\ee
while if we fix the value of $\psi^{(2)}$ on the boundary, we need also an analog of a Gibbons-Hawking term,
\be
S_2 = -\left. \int d^3x \sqrt{-g_\infty} \, (2 \psi \, n^\mu  \partial_\mu \psi + \psi^2/L) \right|_{r=r_\infty} \,.
\ee

The sum $\tilde S_E = S_E + S_ \text{c.t.} + S_{1, 2}$ is now finite in the limit $r_\infty \to \infty$. 
The regularised action becomes, combining the cases of the two possible
boundary conditions above
\be\label{eq:Eover2}
-T \tilde S_{E}  = \int d^2x \left( \frac{\edensity L^2}{2} + \gamma \psi^{(1)} \psi^{(2)}  \right) = \frac{E L^2}{2} + \gamma V_2 \psi^{(1)} \psi^{(2)} \,,
\ee
with $\gamma = 2/3$ in the first case and $\gamma = -4/3$ in the second case.
Here we used the fact that the length of the Euclidean thermal circle is $1/T$ and in the
last equality we assumed that the $\psi^{(i)}$ were constant in space.
For our solutions, at least one of the $\psi^{(i)}=0$.
We will see in Section  \ref{sec:critBfield} below that the result (\ref{eq:Eover2})
for the background (equilibrium) configuration matches the general expectation
for a 2+1 CFT that
the grand canonical
potential function is $\Omega = - E/2$.

\setcounter{equation}{0}
\section{Conductivities}
\label{sec:conductivity}

In this section we will study transport phenomena in our
holographic superconductors.\footnote{%
 We set $L=1$ for simplicity in the remainder of this paper. This choice only
 affects the overall normalisation of the bulk action.
}
In particular, we obtain the electric,
thermal and thermoelectric conductivities as a function of
frequency. Transport describes the response of the system to small external
sources. Therefore, we will need to compute the retarded Greens functions
for the electric and heat currents. In the AdS/CFT correspondence,
these correlation functions are computed by looking at the linear response of the
system to fluctuations of the fields $A_x$ and
$g_{tx}$ in the bulk. These fluctuations are dual to the electric current $J^x$
and energy current $T^{tx}$ operators in the CFT.
At zero spatial momentum, these two fluctuations do not source
any other modes of the metric, Maxwell field or scalar field.
This decoupling simplifies considerably the computation, so we shall
restrict to zero spatial momentum in this paper.

\subsection{Formulae for the conductivities}

Assuming a time dependence of the form $e^{- i \w t}$,
we linearise the Maxwell and Einstein equations above to yield
equations governing perturbations of $A_x$ and $g_{tx}$.
We start with Maxwell's equation
\be
A_x'' + \left[\frac{g'}{g} - \frac{\chi'}{2} \right] A_x'
+ \left[\frac{\w^2}{g^2}  e^{\chi} - \frac{2 \charge^2
\psi^2}{g} \right] A_x  =  
\frac{\phi'}{g} e^\chi \left( -g_{tx}' + \frac{2}{r} g_{tx} \right)  \ .
 \label{eq:axprelim}
\ee
We can caricature this equation in a way that makes the physics of the conductivity much clearer.  If we assume that the radial dependence is not essential, then this equation describes a photon of mass proportional to $q^2 \psi^2$ coupled to a metric fluctuation in $g_{tx}$.  Furthermore, in the probe limit, we assume the gauge field does not back react on the metric, and we set $g_{tx}=0$.  In this limit, we have at least morally the Higgs mechanism and thus expect the charge current response to a magnetic field and the infinite DC conductivity typical of a superconductor.
If we move away from the probe limit and allow $g_{tx}$ to fluctuate, we find remarkably that $g_{tx}$ is governed by only a first order differential equation in the case of zero spatial momentum:
\be
g_{tx}' - \frac{2}{r} g_{tx} + \phi' A_x  =  0 \,. \label{eq:gx} 
\ee
We can substitute this Einstein equation into (\ref{eq:axprelim}) to find
\be
A_x'' + \left[\frac{g'}{g} - \frac{\chi'}{2} \right] A_x'
+ \left[\left(\frac{\w^2}{g^2} - \frac{\phi'^2}{g} \right) e^{\chi} - \frac{2 \charge^2
\psi^2}{g} \right] A_x  =  0 \,. \label{eq:ax}
\ee
The effect of allowing the metric to fluctuate has been to add another contribution to the effective mass that scales as $\phi'^2$.  This extra contribution is the result of restoring translation invariance and will lead to the infinite DC conductivity typical of translationally invariant charged media.\footnote{%
 From a quasiparticle point of view, the charged particles of the system are accelerated by the external
 field.  There may be scattering events, but translation invariance and a net charge means 
 the end result will be acceleration of the entire system.
} 
However, we do not expect this extra mass to affect the response of the system to magnetic fields.  In support of this second claim, we find that if we look at fluctuations with nonzero spatial momentum, in addition to having to consider many more modes of the metric and gauge field, $g_{tx}$ will satisfy a second order differential equation. Therefore we would not be able to simply eliminate $g_{tx}$ from the equation of motion for $A_x$.

These fluctuation equations (\ref{eq:gx}) and (\ref{eq:ax}) 
are solved at
the linearised level by using the equations for the background
in section \ref{sec:phases}.
The asymptotic large $r$ behaviour of the perturbations is
\be\label{eq:agasymptotics}
A_x = A_x^{(0)} + \frac{A_x^{(1)}}{r} + \cdots \,, \qquad g_{tx} =
r^2 g_{tx}^{(0)} + \frac{g_{tx}^{(1)}}{r} + \cdots \,.
\ee
As is standard in AdS/CFT, the leading term determines
a source in the dual theory, while the `normalisable' term will give
the expectation value of the dual current. We shall see this explicitly
shortly.

Preparatory to calculating two-point correlation functions of the currents,
we evaluate the quadratic action for perturbations about a
 solution to the equations of motion.  The on shell (Lorentzian) action is
\be
S_\text{o.s.} \equiv \int_{r_+}^{r_\infty} dr \int d^3 x \, \sqrt{-g_B} {\mathcal L} \ .
\ee
 Expanding the
action to quadratic order in  $A_x$ and $g_{tx}$, and using the equations
of motion (\ref{eq:gx}) and (\ref{eq:ax}), we find that the quadratic action reduces to a surface term:
\be
 S_\text{o.s.}^{(2)} =
\left. \int d^3 x  \, e^{\chi/2} \left(  -
\frac{g}{2} e^{-\chi} A_x A_x' -
g_{tx} g_{tx}'
+ \frac{1}{2} \left(\frac{g'}{g} - \chi' \right) g_{tx}^2
\right) \right|_{r=r_\infty} \ .
\ee
There is no contribution from the lower limit of the radial integral since $g$ and $g_{tx}$ both vanish at the horizon $r=r_+$.
 We would like eventually to take the upper limit of the radial integral to correspond to the conformal boundary of our asymptotically AdS space, $r_\infty \to \infty$.  As usual, $S_\text{o.s.}$ is not finite in this limit and must be regulated by
counter terms.

The counter terms we need to regulate $S_\text{o.s.}$ are the same as those used in section (3.4).    Adding the  appropriate terms together we find that, evaluated on a solution, the regularised quadratic action becomes
\bea \label{osone}
S^{(2)}_1 & = & \lim_{r_\infty \to \infty} \left( S_\text{o.s.}^{(2)} + S_\text{c.t.} + S_1 \right)  \nonumber \\
&= & 
\int d^3x \left(  
\frac{1}{2} A_x^{(0)} A_x^{(1)}  - 3 g_{tx}^{(0)} g_{tx}^{(1)}  -  \left(\frac{\edensity}{2}  - \frac{1}{3} \psi^{(1)} \psi^{(2)} \right)g_{tx}^{(0)} g_{tx}^{(0)}
\right)
\ ,
\eea
and
\bea \label{os2}
S^{(2)}_2 & = & \lim_{r_\infty \to \infty} \left( S_\text{o.s.}^{(2)}  + S_\text{c.t.} + S_2 \right) \nonumber \\
&=&  \int d^3x \left(
\frac{1}{2} A_x^{(0)} A_x^{(1)}  - 3 g_{tx}^{(0)} g_{tx}^{(1)} -  \left(\frac{\edensity}{2}  + \frac{2}{3} \psi^{(1)} \psi^{(2)} \right)g_{tx}^{(0)} g_{tx}^{(0)}
\right)
\ .
\eea
These regulated expressions are now manifestly finite in the limit $r_\infty \to \infty$. 

Equipped with the quadratic action we can obtain the conductivities as follows.
Firstly note that the mixing of the Maxwell perturbation with the metric mode means
that we must consider thermal and electric transport jointly.
These phenomena are described by the matrix of conductivities
\be\label{eq:currents}
\left( \begin{array}{c}
  J_x \\
  Q_x \\
\end{array}\right)
= \left( \begin{array}{cc}
  \sigma & \alpha T \\
  \alpha T & \bar \kappa T \\
\end{array}\right)
\left(
\begin{array}{c}
  E_x \\
  - (\nabla_x T)/T \\
\end{array}
\right) \ .
\ee
Here $J_x$ is the electric current and $Q_x=T_{tx}-\mu J_x$ is the heat
current. We will have all currents moving in the $x$
direction and all sources pointing in that direction.
The electrical conductivity is $\sigma$, the
thermoelectric conductivity is $\alpha$ and the thermal
conductivity is $\bar
\kappa$. The external fields are an electric field $E_x$ and a
thermal gradient $\nabla_x T$. Strictly, the currents are
expectation values. We will drop the angled brackets for
notational convenience. The matrix appearing in
(\ref{eq:currents}) is symmetric due to time reversal invariance.

At a nonzero momentum there will also be mixing with the
condensate, which will result in more terms in
(\ref{eq:currents}). We shall not consider these effects here.

We can solve the equation for the metric perturbation
(\ref{eq:gx}) to obtain
\be\label{eq:gsol}
g_{tx} = r^2 \left(g_{tx}^{(0)} + \int_r^\infty \frac{\phi'
A_x}{r^2} dr \right) \,.
\ee
This shows that $g_{tx}^{(0)}$ is independent of the Maxwell
perturbation, whereas $g_{tx}^{(1)}$ is completely determined by
$A_x^{(0)}$. In fact, by expanding (\ref{eq:gsol}) at large radius and
using (\ref{eq:phiasymptotic}) and (\ref{eq:agasymptotics}) we have
\be
g_{tx}^{(1)} = \frac{\rho}{3} A_x^{(0)} \,.
\ee

Given a solution to the equations, we compute the currents by
differentiating the action with respect to the boundary values of
the dual bulk fields
\be\label{eq:diff}
J_x = \frac{\delta S^{(2)}}{\delta A_x^{(0)}} \,, \qquad T_{tx} = \frac{\delta S^{(2)}}{\delta g_{tx}^{(0)}} \,.
\ee
In general we can study these currents in the presence of
arbitrary sources $A_x^{(0)}$ and $g_{tx}^{(0)}$. Firstly let us
set $g_{tx}^{(0)} = 0$. Thus there is no source for heat flow, so
that $\nabla_x T = 0$. In this case, the electric and heat currents are given by
\be
J_x = A_x^{(1)} \,, \qquad Q_x = - 3 g_{tx}^{(1)} - \mu A_x^{(1)} = -
\rho A_x^{(0)} - \mu A_x^{(1)} \,.
\ee
In performing the differentiation, we can note that because we are looking
at linearised equations, $A_x^{(1)}$ will be proportional to $A_x^{(0)}$.

Setting $\nabla_x T = 0$ in (\ref{eq:currents}) we can therefore read off the electrical conductivity
\be
\sigma = \frac{J_x}{E_x} = \frac{- i A_x^{(1)}}{\w A_x^{(0)}} \,,
\ee
and the thermoelectric conductivity
\be\label{eq:alpha}
T \alpha = \frac{Q_x}{E_x} = \frac{i \rho}{\w} - \mu \sigma \,.
\ee
In these expressions we used the fact that $A_x^{(0)}$ is the boundary background
potential, and that $E_x = - \pa_t A_x^{(0)}$. The relation (\ref{eq:alpha}) between
thermoelectric and electrical conductivities is the same as that found in 
\cite{Hartnoll:2007ip}, in the absence of a charged condensate. The simple relationship between
electric and thermoelectric conductivities means that we can focus on computing the electric
conductivity in the remainder of the paper.

To obtain the thermal conductivity, we now set $A_x^{(0)}=0$. Using the fact that our
backgrounds have either $\psi^{(1)}=0$ or $\psi^{(2)}=0$, it follows from (\ref{eq:diff})
that
\be
Q_x = - \edensity g_{tx}^{(0)}.
\ee
Now we need to relate $g_{tx}^{(0)}$ to a thermal gradient. This is a straightforward
computation.  (See for instance the appendix of \cite{Hartnoll:2008hs}.) One obtains
\be
g_{tx}^{(0)} = - \frac{\nabla_x T}{i \w T} \,.
\ee
Combining the last two equations and (\ref{eq:currents}) with $E_x = 0$ gives
\be\label{eq:kappa}
\bar \kappa = \frac{i \edensity}{\w T} \,.
\ee
The divergence as $\w \to 0$ is exactly what we should expect from a translationally
invariant system. Conservation of momentum means that a DC energy current, which is
also a momentum, cannot relax.

\subsection{Numerical results for the conductivity}

\begin{figure}[h]
\centerline{a) \epsfig{figure=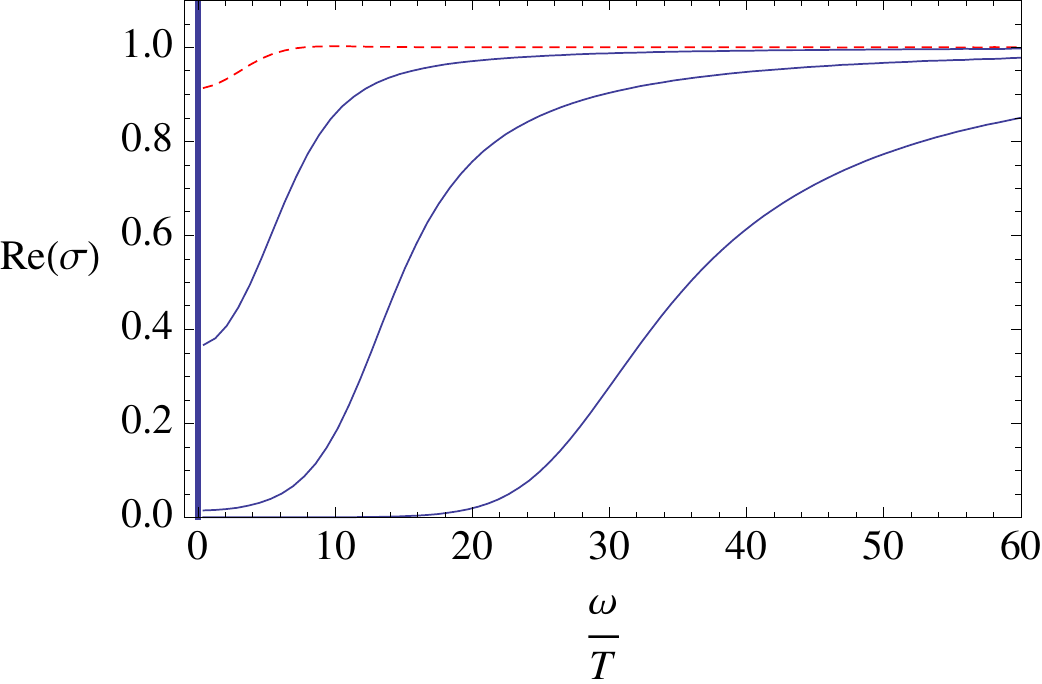, width=3in}  b) \epsfig{figure=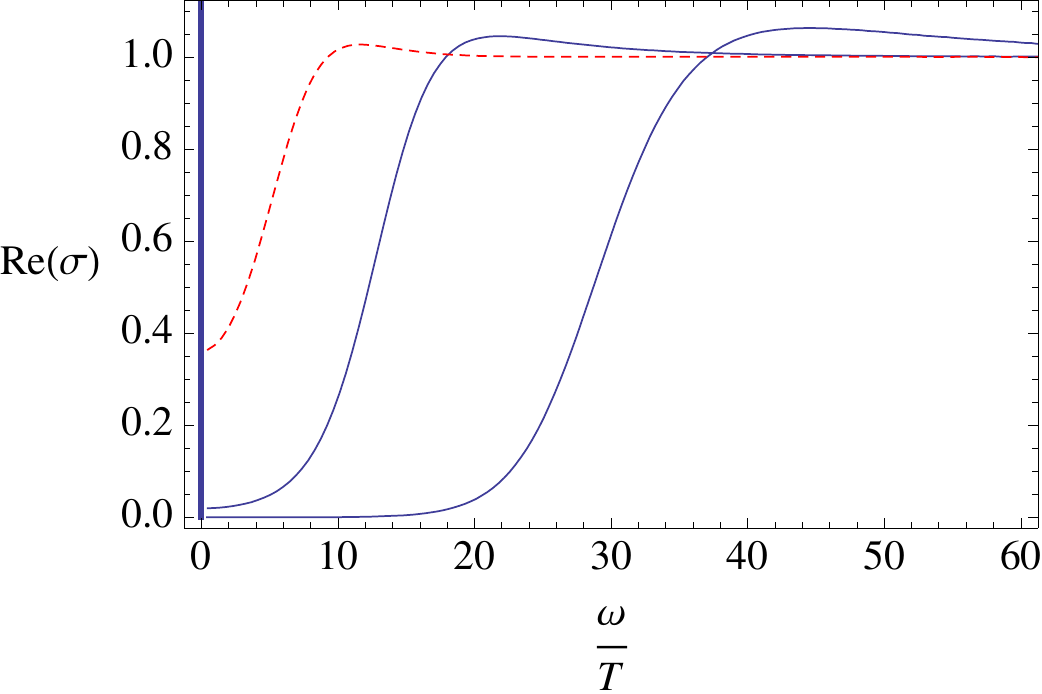, width=3in}}
\caption{
\label{realsliceplots}
The dashed red line is the real part of the conductivity at
$T=T_c$  (for $q=3$).  The blue lines are the same conductivities at
successively lower temperature: a) The dimension one operator with
$T/T_c = 0.810 $, 0.455 and 0.201; b) the dimension two operator
with
$T/T_c = 0.651$ and 0.304. There is a delta function at the origin in all
cases.
 }
\end{figure}

\begin{figure}[h]
\centerline{a) \epsfig{figure=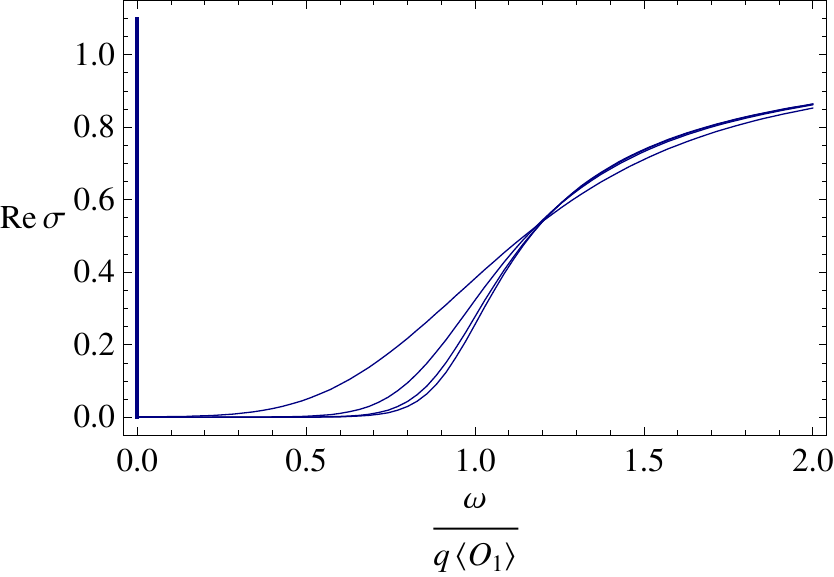, width=3in}  b) \epsfig{figure=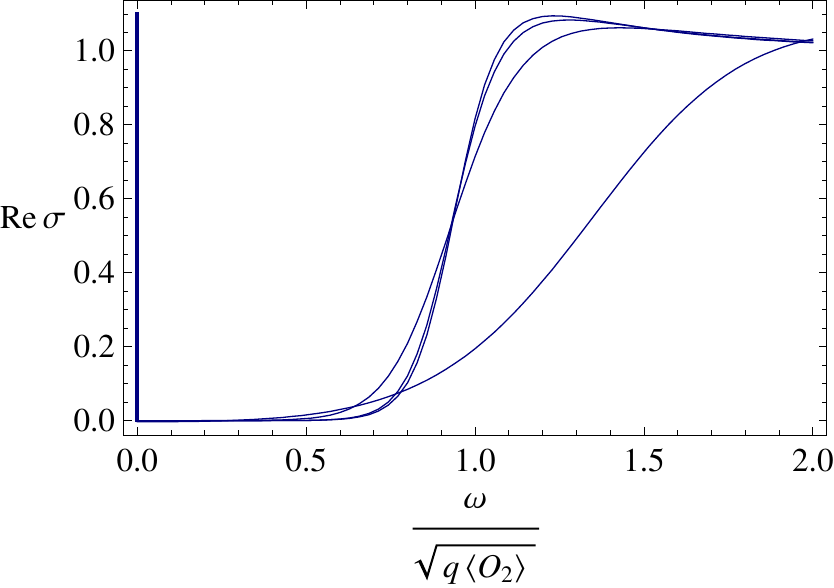, width=3in}}
\caption{
\label{lowtempplots}
We plot the real part of the conductivity as a function frequency normalized by the condensate, either
$ \charge \langle {\mathcal O}_1 \rangle$ or $ \sqrt{ \charge \langle {\mathcal O}_2 \rangle}$ as appropriate.  This data was taken at low temperature, $T = 0.03 \  \charge \langle {\mathcal O}_1 \rangle$  and $T = 0.03  \sqrt{ \charge \langle {\mathcal O}_2 \rangle}$ for a variety of charges $\charge = 1, 3, 6$ and $12$.  The curves with steeper slope correspond to larger $\charge$. There is a delta function at the origin.
 }
\end{figure}

We make several observations about 
figures \ref{realsliceplots} and \ref{lowtempplots} which display our numerical results for the conductivity as a function of frequency $\omega$.  To understand these plots, it is useful to put them in the context of conductivity results from previous work \cite{Herzog:2007ij}, \cite{Hartnoll:2007ip}, \cite{Hartnoll:2008hs}, and \cite{Hartnoll:2008vx}.

From the earliest of these papers \cite{Herzog:2007ij}, we know that the conductivity at vanishing charge density and vanishing condensate is a constant independent of $\omega$.  This independence
can be understood as a consequence of classical electromagnetic self-duality of the dual gravitational theory at the quadratic level.  From the next paper \cite{Hartnoll:2007ip} in this sequence, we have results for the conductivity as a function of charge density in the absence of a scalar condensate.  If we work in the limit where the charge density is small compared to the temperature, we recover the frequency
independent result of \cite{Herzog:2007ij}, but in general the dependence on $\omega$ is more complicated.  In particular, we see the minimum in $\mbox{Re}(\sigma)$ at $\omega=0$ displayed by the dashed curves in figure \ref{realsliceplots}.  The $\mbox{Im}(\sigma)$, not plotted, has a pole at $\omega=0$.  From the Kramers-Kronig relations, which follow from causality, one concludes that the real part of the conductivity has a Dirac delta function at $\omega=0$  that is invisible to the numerics because of its infinitesimal width. 
Recall that one of the Kramers-Kronig relations is
\be
\mbox{Im}[\sigma(\omega)] = -\frac{1}{\pi} {\mathcal P} \int_{-\infty}^\infty \frac{\mbox{Re}[\sigma(\omega')] d\omega'}{\omega'-\omega} \ .
\ee
From this formula we can see that the real part of the conductivity contains a delta function $\mbox{Re}[\sigma(\omega)] = \pi \delta(\omega)$, if and only if the imaginary part has a pole, $\mbox{Im}[\sigma(\omega)] = 1/\omega$. 
 There is also a Ferrell-Glover-Tinkham sum rule which follows from similar arguments and which states that $\int \mbox{Re}(\sigma) d\omega$ is a constant independent of the temperature.  Thus the dip in the real part of the conductivity at $\omega=0$ is related to the residue of the pole in the $\mbox{Im}(\sigma)$ and strength of the Dirac delta function in the $\mbox{Re}(\sigma)$.

This Dirac delta function for $T>T_c$ is naively surprising because it implies an infinite DC conductivity in the normal phase.  This infinite conductivity is not superconductivity and results instead from translation invariance. A translationally invariant, charged system does not have a finite DC conductivity because application of an electric field will cause uniform acceleration.  If we were to break this translation invariance by for example introducing impurities, the delta function at $\omega=0$ would acquire a width for $T>T_c$ and the conductivity would become finite. This effect of impurities was studied in an AdS/CFT setting in \cite{Hartnoll:2008hs}.  

In our previous paper \cite{Hartnoll:2008vx}, we did not see this infinite conductivity above $T>T_c$.  The reason we did not see it is that we worked in a probe limit where the gravitational background was fixed and the abelian-Higgs sector\footnote{Our bulk matter is like a traditional abelian-Higgs model, but it does not have the usual $\psi^4$ term in the potential.}  decoupled. By fixing the background, we implicitly broke translation invariance. Technically this occurs because the electric and energy
currents decouple, as we discussed at the start of this section.
For $T>T_c$ we had a pure Schwarzschild-AdS background and thus recovered the frequency independent conductivity of \cite{Herzog:2007ij}.
It was only for $T<T_c$ that the $\mbox{Im}(\sigma)$ developed a pole.

With this review of previous results, the structure of figure \ref{realsliceplots} should be clear.
We see from the dashed curves in figure \ref{realsliceplots} that $\mbox{Re}(\sigma)$ has a minimum at $\omega=0$ for $T = T_c$ and indeed, although not plotted, also for $T>T_c$ since the background becomes the electrically charged black hole studied in \cite{Hartnoll:2007ip}.  If we were to plot $\mbox{Im}(\sigma)$, we would see a pole in $\omega=0$, and by the Kramers-Kronig relations would conclude that there is also a Dirac delta function in $\mbox{Re}(\sigma)$.  In other words, we have infinite DC conductivity for temperatures above $T_c$.  

For $T<T_c$, this minimum in the $\mbox{Re}(\sigma)$ at $\omega=0$ becomes increasingly pronounced and eventually develops into a gap, as was seen in \cite{Hartnoll:2008vx} in the probe limit.
For lower $q$ the low temperature gap becomes less pronounced, see figure \ref{lowtempplots}. As far
as the accuracy of our numerics permits, the conductivity still appears to vanish at zero temperature
over a finite range of small frequencies. The residue of the pole in $\mbox{Im}(\sigma)$ and the strength of the Dirac delta function become much larger as well.  There is an additional contribution to the strength of the Dirac delta function coming from condensation of the scalar.   Note that the low temperature plot of figure \ref{lowtempplots} gives strong evidence that for relatively large values of the charge of the scalar field $q \gtrsim 3$, the size of the zero temperature gap, $\w_g$, can be associated in the dimension one case with $q \langle {\mathcal O}_1 \rangle$ and in the dimension two case with $\sqrt{q \langle {\mathcal O}_2 \rangle}$.
Let us take this observation as the definition of $\w_g$. Thus the vertical axis in our figure \ref{condensate} can be thought of as $\w_g/T_c$.

The pole in the conductivity for $T < T_c$ is (partly) due to different physics than that for $T > T_c$. It is
not exclusively due to translational invariance and would persist in the presence of impurities. It is instead closely tied to the spontaneous breaking of the $U(1)$ symmetry and the expulsion of magnetic fields. In sections \ref{sec:london} and \ref{sec:mass} below, we shall discuss the physics of magnetic fields and Higgsing in this model in greater detail.

As we lower $T$ past $T_c$ we therefore expect to see a non-analytic change in the strength of the delta function, reflecting the onset of superconductivity. Indeed,
there is a jump in the derivative of the strength of the delta function with respect to temperature, indicative of a second order phase transition, as we now discuss.
 Above the transition temperature, we know from ref.\ \cite{Hartnoll:2007ip} that as $\w \to 0$ the normal phase conductivity satisfies\footnote{This pole vanishes in the probe limit since that limit requires $\rho \propto 1/q$ as $q \to \infty$.}
\be
\mbox{Im}(\sigma_n) = \frac{4 \rho^2 }{3(4r_+^4+\rho^2)} \frac{r_+}{\omega}+ {\mathcal O}(1)
\; \; \; \; \mbox{where} \; \; \; \; T = \frac{12 r_+^4 - \rho^2}{16 \pi r_+^3} \ .
\ee
From our numerics, we can study this pole for $T<T_c$. We find that  there is an additional contribution to the pole below the critical temperature. Close to $T_c$, it takes the form
\be
\mbox{Im}(\sigma)  = \mbox{Im}(\sigma_n) +  \frac{A}{\omega} (T_c -T) \ .
\ee
The coefficient $A$ depends on the charge of the scalar field and its boundary condition at infinity. Consider, for example, the case where  $\charge=3$.
For the dimension one operator, $A \approx 15$ and for the dimension two operator,  $A \approx 12$. Therefore  $\pa \text{Im}(\sigma)/\pa T$ is discontinuous across $T_c$. In computing these
slopes, $\rho$ is held fixed as $T$ is varied.

Given that we appear to see an exact gap emerge as $T \to 0$, i.e. that $\text{Re}(\sigma)$ vanishes identically for $\w < \w_g$ at $T=0$, we can expect that at small but finite temperatures, the zero frequency limit of the real part of the conductivity is governed by thermal
fluctuations
\be\label{eq:exponential}
\lim_{\omega \to 0} \mbox{Re} (\sigma)\sim e^{- \Delta_i/T} \
\ee
when $\Delta_i / T \gg 1$. We can further ask whether $\Delta_i$ is related to $\w_g$. If we parametrise this relation by $\Delta_i = \alpha_i \w_g$ then,
numerically using the lowest temperatures accessible to us, we find that
\be
\begin{array}{c|cc}
\charge & \alpha_1 & \alpha_2 \\
\hline
3 & 0.13 & 0.14 \\
6 & 0.34 & 0.30 \\
12 & 0.45 & 0.44 \\
\end{array}\label{eq:table}
\ee
From ref.\  \cite{Hartnoll:2008vx}, we know that in the probe limit, which corresponds to large $\charge$,
these values of $\alpha_i$ should become close to 1/2. Extracting these values numerically
is delicate as one needs to obtain very low temperatures. The accuracy of our numerics
decreases with $q$ and we have not quoted the $q=1$ values because our numerical results are insufficiently robust. On the other hand, the exponential
behaviour (\ref{eq:exponential}) is clearly seen, as is the fact that the $\alpha_i$ are less than $1/2$
for finite $q$.

\subsection{Comments on gaps and pairing}

The fact that the conductivity in the superconducting phase tends toward the value of the
normal phase conductivity as $\w \to \infty$ indicates that the degrees of
freedom responsible for high frequency conductivity are those of the normal phase.

At zero temperature and frequencies $\w \lesssim \w_g$ conduction is non-dissipative, i.e.\ the real part of the conductivity vanishes identically, at least within the accuracy of our numerics.\footnote{%
Recall we defined $\w_g$ to be $q \langle \ocal_1 \rangle$ or $\sqrt{q \langle \ocal_2 \rangle}$, motivated by figure \ref{lowtempplots}.
}
The region where the conductivity vanishes implies a corresponding gap in the charge spectrum ---
if there were asymptotic charged states with energy below $\w_g$, they would contribute to the conductivity at these low frequencies. Yet this statement is not entirely true: we know that there is a Goldstone boson in our symmetry broken phase. Usually a Goldstone boson leads to a nonzero conductivity all the way down to $\w=0$ due to the presence of multi-Goldstone boson states. The fact that this does not occur in our models is possibly a large $N$ effect.

In a standard weak coupling picture of superconductivity, the gap $\w_g$ is understood as the energy required to break a Cooper pair into its constitutive electrons. The energy of the constituent quasiparticles is given by $\Delta$ and then $\w_g$ would be some integer multiple of this energy. The non-integer relation between $\Delta$ and $\w_g$ in table (\ref{eq:table}) shows that we are clearly not in a weak coupling regime and that  such a quasiparticle picture is not applicable, except perhaps in the $q \to \infty$ limit, in which we recover the probe result \cite{Hartnoll:2008vx} $\w_g = 2 \Delta$.
At strong coupling we would expect to be closer to a Bose-Einstein condensate scenario than
to a BCS like weak pairing description.
On the other hand, a non-integer relation is almost inevitable if $\w_g/\Delta$ is to depend continuously
on the charge $q$. It may be that microscopic realisations of holographic superconductors, that is, embeddings of our setup into string theory, will place constraints on the masses and charges of the scalars that condense.

It is interesting to note that for the dimension two condensate and charges $q \gtrsim 3$, even though $\alpha_i$ changes by at least a factor of three,  $\w_g/T_c$ remains close to the value $8$. The constancy of $\w_g/T_c$ can be seen by comparing Fig. 4b and Fig. 1b. We conclude that the energy to `break apart' the condensate is insensitive to the charge for $q \gtrsim 3$. In other words, the probe limit is reached rapidly for this ratio.

\setcounter{equation}{0}
\section{Critical magnetic fields}
\label{sec:critBfield}

In this section, we start our investigation of  the effect of magnetic fields on our holographic 
superconductor. We will argue that  our model  behaves as a type II superconductor.  Recall that the difference between type I and type II superconductors lies in the way
the Meissner effect disappears as the temperature of the material is raised, as we now review.

The Meissner effect is the observation that at low temperatures superconductors expel magnetic field lines.  The existence of this effect implies that there is a critical magnetic field $H_c$ above which the superconducting order is destroyed and the material reverts to its normal state: The superconductor must perform an amount of work $H^2 V_3 / 8 \pi$ to expel an applied field $H$ from a volume $V_3$.  The critical field strength is then obtained by equating this work to the difference in free energies between the normal and superconducting states of the material.  For $H > H_c$ it is no longer
thermodynamically favorable to superconduct.

How the superconductor reverts to a normal state as the magnetic field is increased depends on the nature of the material.  For type I superconductors, there is a first order phase transition at $H=H_c$, above which magnetic field lines penetrate uniformly and the material no longer superconducts.  For  type II superconductors, vortices start to form at $H=H_{c1}$.  In the vortex core, the material reverts to its normal state and magnetic field lines are allowed to penetrate.  The vortices become more dense as the magnetic field is increased, and at an upper critical field strength $H=H_{c2}$, the material ceases to superconduct.  

In determining whether our model is type I or type II, we are faced with the limitation that currents in the model do not source electromagnetic fields. One immediate consequence of this limitation is notational.  As the material does not produce its own magnetic fields, the applied magnetic field is the actual magnetic field and we can set $H=B$.
Given the limitation, the Meissner effect cannot strictly speaking exist: to exclude the magnetic field, the current produced by the external magnetic field must produce an equal and opposite canceling field inside the sample.  
However, we will show in a later section that holographic superconductors do generate the currents required to expel magnetic fields
(the London equation) and that the theory can consistently be weakly gauged. Therefore
we assume for the moment that (a gauged version of) our model attempts to expel fields in the usual way for superconductors.

This brings us to a subtlety. We would like to work with a 2+1 dimensional model interacting with a 3+1 dimensional electromagnetic field.  This is a realistic setup for a thin film superconductor.
We apply the magnetic field normal to the material.  Assume
for the moment that the 2+1 dimensional sample is a disk of radius $R$.
In order for the disk to expel the magnetic field, the disk must produce a current circulating around the perimeter.  Solving Maxwell's equations, this current will expel a field not only in the area $\pi R^2$ of the disk but in a larger volume of size $V_3 \sim R^3$.  As mentioned before, 
the amount of work that the superconductor must do to 
exclude an applied magnetic field from a volume $V_3$ scales as $H^2 V_3$.  In the large $R$ (thermodynamic) limit, the superconductor does not have enough free energy available to expel a magnetic field from such a large region; the difference in the free energies between the normal and superconducting phases can scale extensively only as $R^2$.  Thus magnetic fields of any non-vanishing strength will penetrate a thin superconducting film and $B_c=0$. This argument is illustrated in figure \ref{fig:ExpelFlux} below.

\begin{figure}[h]
\centerline{ \epsfig{figure=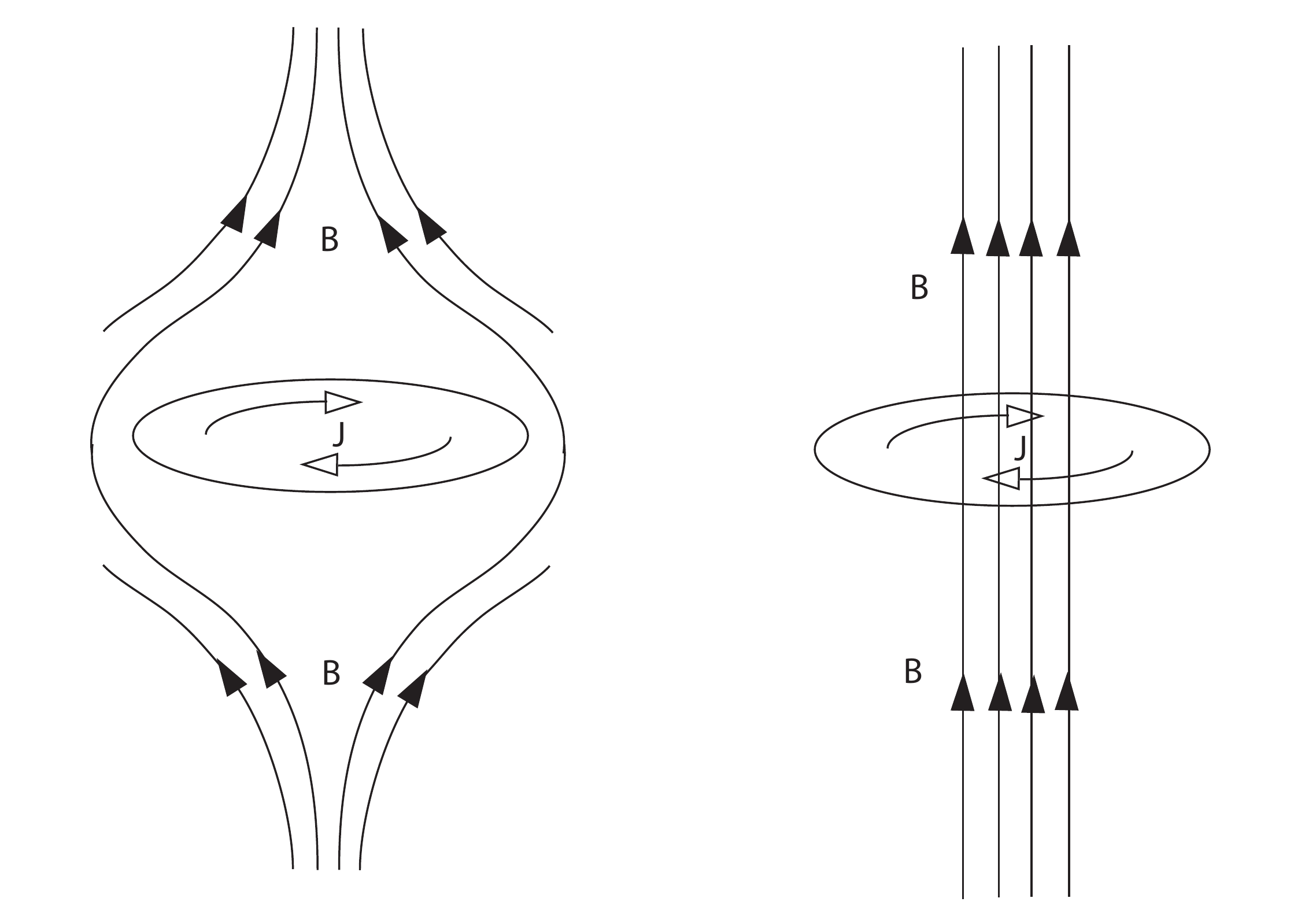, width=3.5in} }
\caption{
\label{fig:ExpelFlux}
In order to prevent the flux from penetrating the superconductor, of area $R^2$, the currents would have to do enough work to expel the field from a volume of size $R^3$, as shown in the left hand figure. This work cannot be supplied by the free energy gain of superconducting on the thin film. Therefore, the flux always penetrates the film, as shown in the right hand figure.
 }
\end{figure}

So far, our discussion applies to any thin superconductor in a perpendicular magnetic field.
For our model to be type II, we still need to establish that the material remains a superconductor for values of the magnetic field up to some upper critical field strength $B_{c2}$.  To that end, we now investigate the largest value of the magnetic field for which our scalar condenses.

\subsection{Superconducting droplets}
\label{sec:Binstab}

In this section, we will examine the effect of a constant background magnetic field on the dynamical ability of the condensate to form.\footnote{The computations in this subsection are similar to those in \cite{Albash:2008eh}, but our interpretation is different.} For each temperature $T < T_c$, we expect there to be some critical field strength $B_{c2}$ above which the condensate cannot form. We find $B_{c2}$ by starting in a phase with a large magnetic field and no condensate, and showing that this phase develops an instability towards condensation of the scalar as $B$ is lowered.

Our starting point is a dyonic black hole background. This is a Reissner-Nordstrom AdS black hole
with both electric and magnetic charges and no scalar hair. As we have already seen, the electric charge of the black
hole gives the charge density of the field theory. The magnetic charge gives the value of the background magnetic field, as explained in, for instance \cite{Hartnoll:2007ai}. The form of the solution is well known (see e.g. \cite{Hartnoll:2007ai}). The metric takes the form (\ref{eq:metric}) with $\chi=0$ and
\be\label{eq:dyonic}
g(r) = r^2 - \frac{1}{4r r_+} \left(4 r_+^4 + \rho^2 +B^2\right) + \frac{1}{4r^2}(\rho^2 + B^2)  \ .
\ee
In this section, we find it convenient to work in polar coordinates $dx^2 + dy^2 = du^2 + u^2 d\varphi^2$.
To the vector potential (\ref{eq:Aandpsi}) we have to add a magnetic component:
\be
A = \rho \left( \frac{1}{r_+}- \frac{1}{r} \right)\, dt + {1\over 2} Bu^2 d\vp .
\ee
The horizon radius $r_+$ is determined implicitly by the temperature via
\be\label{eq:dyonictemp}
T = \frac{12 r_+^4 - \rho^2 - B^2}{16 \pi r_+^3 } \ .
\ee
We will choose a gauge in which the scalar $\psi $ is real.
Given the above background, we will treat the scalar field as a perturbation and look for a solution that is well behaved at the horizon (no logarithmic divergence) and chooses one of the two fall-offs, $\psi^{(1)}=0$ or $\psi^{(2)}=0$ at the boundary, corresponding to our two choices of boundary operator.  We interpret the existence of such a zero mode solution as the onset of an instability for the condensate of the corresponding scalar operator to form.

We are interested in static, axisymmetric solutions in which all fields are independent of $t$ and $\vp$.
The scalar field thus satisfies the equation
\be\label{eq:psieq}
{1\over u}\pa_u(u\pa_u\psi) + \pa_r(r^2g\pa_r\psi) + \left[{\charge^2 \rho^2 \over g r_+^2} (r-r_+)^2 - {\charge^2 u^2 \over 4 } B^2  + 2r^2 \right] \psi = 0 \ .
\ee
This linear PDE can be solved by separation of variables:
\be\label{psisep}
\psi(r,u) = R(r) U(u)
\ee
where $U(u)$
solves the equation for a two dimensional harmonic oscillator with frequency determined by $B$:
\be
U^{''} + {1\over u} U' - \left( \frac{quB}{2} \right)^2 U = - \lambda U \ ,
\label{eq:Ueq}
\ee
$R$ satisfies
\be
(r^2 g R')' + \left[{\charge^2 \rho^2 (r-r_+)^2\over g r_+^2} + 2r^2\right] R = \lambda R \ ,
\label{eq:Req}
\ee
 and the separation constant $\lambda = \charge n B$. Clearly, the condensate is now clumped, with a Gaussian profile.
 We expect that the lowest mode $n=1$ will be the first to condense and lead to the most stable solution after condensing. Therefore we choose
 \be\label{eq:gaussian}
 U(u) = \exp(-\charge Bu^2/4) \,.
 \ee

\begin{figure}[h]
\centerline{a) \epsfig{figure=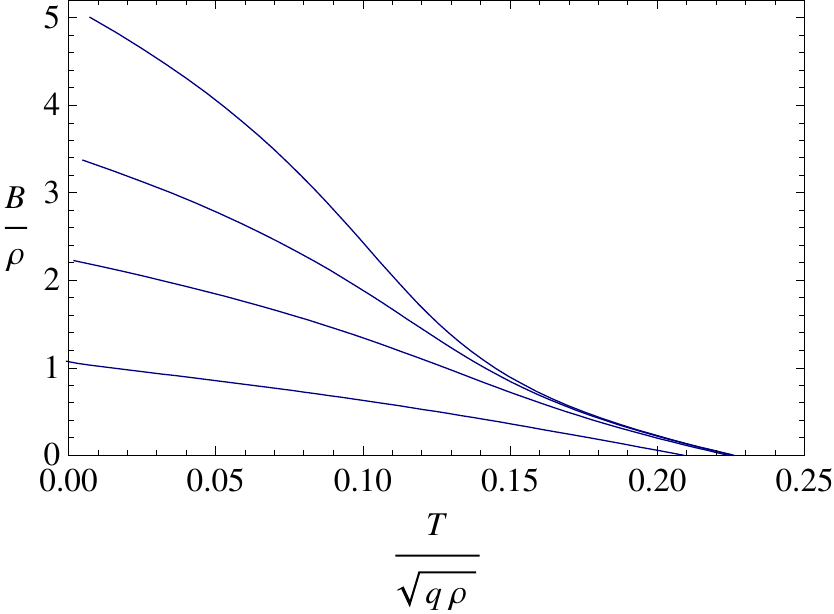, width=3in}  b) \epsfig{figure=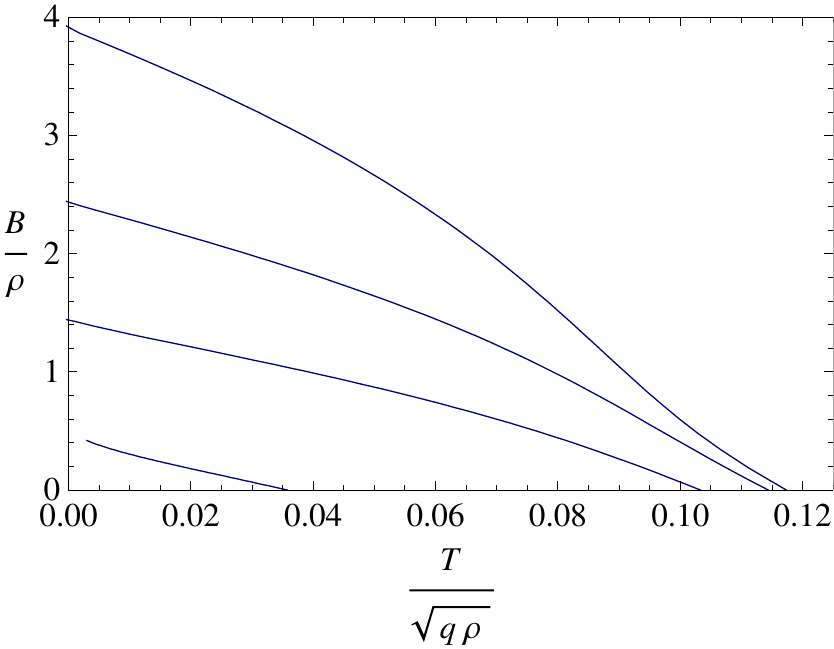, width=3in}}
\caption{
\label{fig:critB}
We plot the critical field $B_{c2}$, below which a droplet of condensate forms, versus temperature: a) The ${\mathcal O}_1$ case with, from right to left, $q=12$, 6, 3, and 1; b) The ${\mathcal O}_2$ case with, from right to left, $q=12$, 6, 3 and 1. In the lower left region there are circular droplets of superconducting condensate. In the top right region there is no superconductivity.
 }
\end{figure}

From the equation of motion (\ref{eq:Req}), it is straighforward to analyze the near boundary and near horizon asymptotics of $R(r)$.  The near boundary behavior is the same as was found above in (\ref{eq:psiasymptotic}).  The near horizon behavior on the other hand takes the form
\be
R = c_0 + c_1 \ln (r/r_+ - 1) + \ldots
\ee
We are looking for a regular zero mode and thus set $c_1=0$.  The equation for $R$ is linear and so $c_0$ is arbitrary and can be set to one.  Requiring either $\psi^{(1)}=0$ or $\psi^{(2)}=0$ at the boundary thus produces a curve of solutions in the $(\rho,B)$ plane.  We call this curve $B_{c2}(\rho)$ and have plotted it in figure \ref{fig:critB}. There will be a localised droplet of condensate on the lower left region of these figures.


It is important to note that in order to see the onset of this instability we only needed to work to first order in the condensate, which is small just below the transition. We will see in a later section that at next order the condensate causes magnetisation currents. A proper treatment of these currents would have to include the effect of the backreaction of these currents on the external magnetic field via Maxwell's equations. For instance, at sufficiently low temperatures the superconducting droplets presumably grow and trap the magnetic flux into vortices. Magnetic screening is an important feature of vortex physics, yet
this would take us beyond the AdS/CFT model. It is fortunate, therefore, that these effects are not important at the transition itself.

\subsection{Thermodynamics}

In this section we investigate quantitatively the effect of the magnetic field on the free energy of our model.  We are not able to carry out the most obvious calculation, which is a determination of the magnetic field dependence of the free energy in the superconducting phase, since we don't have the general solution for a hairy black hole with both electric and magnetic charge.  However, we will see clear evidence that the transition is second order at $B=0$, where we can compute the free energy. In passing, we also remark on the strong diamagnetism of our material in the normal phase.

We choose to work in the canonical ensemble, at fixed charge density.  Recall the thermodynamic identity
\be
E + P V = ST + \mu Q \,,
\ee
where $E$ is the total energy, $P$ is the pressure, $V$ the volume,
$S$ the entropy, and $Q$ the total charge.
The combination $-PV$ is also the value of the potential function $\Omega$
in the grand canonical ensemble.  We would like to work instead with the free energy $F$ in the canonical ensemble
\be
F = \Omega + \mu Q = - PV + \mu Q = E - ST \ .
\ee
For our particular theory, we have an additional relation that comes from the tracelessness of the stress-tensor.  However, we have to be a little careful here because in the presence of a magnetic field $B$, it is possible to define two different pressures.   The diagonal spatial components of the stress-tensor are related to $P$ via a magnetization $M=mV$, $T^{ii} = P - m B$.  Tracelessness thus implies
$E = 2 (PV - MB)$, and we may write
\be\label{eq:FBO}
F(B, {\mathcal O}) = -\frac{E}{2} + \mu Q - MB\ .
\ee

First, we would like to compute the value of the free energy of a configuration with
a magnetic field but no condensate. We wrote the corresponding gravity background,
a dyonic black hole, in (\ref{eq:dyonic}) above and the temperature in (\ref{eq:dyonictemp}).
To determine $\Omega$, one calculates the on-shell value of the regulated Euclidean action.  The action is regulated by the usual Gibbons-Hawking term and a boundary cosmological constant.  We do not reproduce the details here as they can be found in \cite{Hartnoll:2007ai}.  The result is
\be
\Omega(B,0) = \frac{1}{r_+} \left(- r_+^4 - \frac{\rho^2}{4} + \frac{3B^2}{4} \right) V\ .
\ee
To determine $F$, we must add $\mu Q = \rho^2 V$:
\be
F(B,0) =  \frac{1}{r_+} \left(- r_+^4 + \frac{3 \rho^2}{4} + \frac{3B^2}{4} \right) V\ .
\label{FBzero}
\ee

The magnetic field dependence of (\ref{FBzero}) is unusual and means that, while not superconducting, the normal phase of the material is strongly diamagnetic at low temperatures.  Note that at small values of $\rho$ and $B$, we may replace $r_+$ with $4 \pi T/3$.  Thus the magnetic susceptibility $\chi = \partial^2 F /  \partial B^2$ becomes of order $1/T$.  The dimensionless quantity $\chi T$ is naively of order one for this model.  If we normalize the action to be consistent with established AdS/CFT dualities, such as the M2-brane theory, then the susceptibility will scale with a power of $N$ in the large $N$ limit.  Compare this result with a typical 3+1 dimensional, non-ferromagnetic metal.  In 3+1 dimensions, the susceptibility is dimensionless, and approximating a metal as a free electron gas, the susceptibility is suppressed by a power of the fine structure constant and is typically tiny. There is a general lesson here: quantum critical theories will often be strongly magnetic because there is no small coupling or scale to suppress the magnetic susceptibility.

The smallness of the susceptibility for a 3+1 dimensional electron gas explains an approximation that is typically made in calculating $H_c$ for superconductors.  Namely, the dependence of the free energy of the normal phase on the magnetic field is neglected.  Because of the large diamagnetism of our model, we clearly would not be able to make this approximation.  Nevertheless, the argument above that $B_c=0$ for our model still holds.  The reason, as we explained above, is that to be a perfect diamagnet, the superconductor in 2+1 dimensions essentially has to have an infinite susceptibility that scales with the system size $R$.

Next we compute the free energy of the system with a condensate and no magnetic field.
The parameter $\edensity$ in the expansion (\ref{eq:gbryexp}) and (\ref{eq:chibryexp}) is naturally interpreted as the energy density in the boundary field theory, $\edensity V = E$.  Meanwhile, we can obtain $\mu$ and $Q = \rho V$ from the asymptotic expansion of $\phi$ (\ref{eq:phiasymptotic}).  Putting the pieces together, we have from (\ref{eq:FBO})
\be
F(0, {\mathcal O}) = \left( -\frac{\edensity}{2}  + \mu \rho \right) V \ .
\label{FzeroO}
\ee
This result agrees with the value of the regulated on-shell Euclidean action calculated in (\ref{eq:Eover2}).  The value of the regulated action is $\Omega = - \edensity V /2$.

\begin{figure}[h]
\centerline{ \epsfig{figure=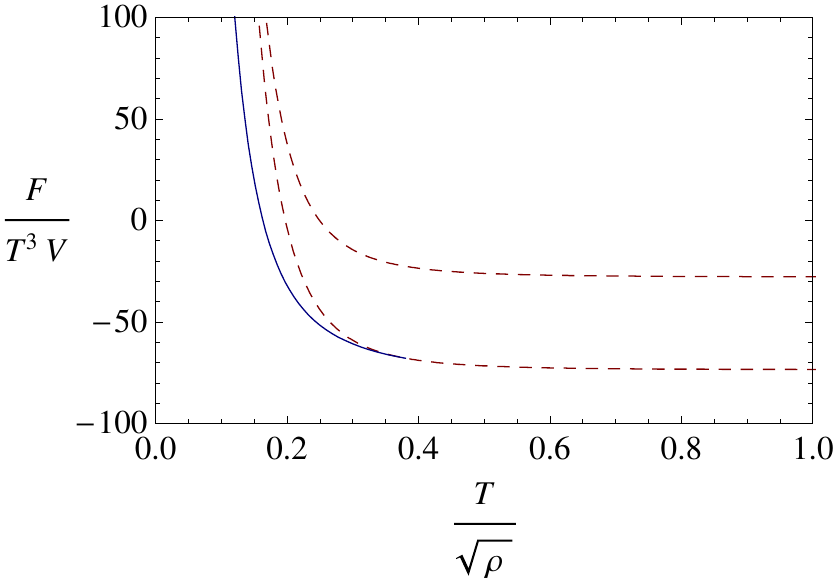, width=4in} }
\caption{
\label{fig:free_energy}
The solid blue line is the free energy of the hairy black hole ($B=0$). The dotted red line tangent to this blue line is the free energy of the electric black hole with no condensate ($B=0$).  The dotted red line at the top is the free energy of a dyonic black hole (no condensate) with $B/T^2 = 20$.  The plot is for ${\mathcal O}_1$ with $\charge = 3$. We see that the magnetic field raises the free energy of the normal phase.
 }
\end{figure}

Being careful to compare (\ref{FBzero}) and (\ref{FzeroO}) at fixed $T$ and $\rho$, figure \ref{fig:free_energy} displays the free energies for representative values of the parameters.   
Note the continuous second order phase transition between the normal and superconducting phases at $B=0$. In order to show the continuity of the transition at finite $B$, we would need
the black hole background with both condensate and magnetic field, at least to second order near the critical temperature. The qualitative similarity between the instability with and without magnetic fields suggests that the transition will be second order in general.



%

Having computed the free energy, it is straightforward to obtain the specific heat by
differentiating: $c = - T/V \, \pa^2 F/\pa T^2$. At low temperatures we do not find the exponential
suppression of the specific heat typical of s wave superconductors. Rather the specific heat vanishes
as a power law as $T \to 0$. It is difficult to determine the precise power due
to numerical sensitivity at low temperatures. A likely source of this power law behavior is
the presence of a Goldstone mode in our system.

A power law rather than exponential behavior at low temperatures is also
observed for the quantity $n_s(T) - n_s(0)$. Here we define $n_s$ to be the
coefficient of the pole in the imaginary part of the conductivity as $\w \to 0$.

\setcounter{equation}{0}
\section{Magnetically induced currents in the superconducting phase}
\label{sec:london}

In this section we continue our study of the superconducting phase in the presence
of a finite magnetic field. The key physics we wish to examine are the currents generated
by the background magnetic field. These currents are responsible for the Meissner
effect once the theory is coupled to dynamical photons.

There will be two main discussions in this section. Firstly, we shall exhibit the London equation
analytically at low temperatures. Secondly, we study the phase diagram of the (ungauged) theory
in the presence of a homogeneous background magnetic field. 

\subsection{The London equation at low temperatures}

We would like to explain how the London equation arises in our model.\footnote{There
is some overlap of this section with \cite{Maeda:2008ir} which
appeared as we were completing this work.}
The London equation\footnote{Recall that we have defined $n_s$ to be the coefficient of the pole at $\omega =0$ in ${\text {Im}}(\sigma)$.}
\be
  J_i(\omega, k)  = - n_s A_i(\omega, k)
\ee
was proposed (in a gauge where the order parameter is real) 
to explain both the infinite conductivity and the Meissner effect of superconductors. This equation is understood to be valid where $\omega$ and $k$ are small compared to the scale at which the system loses its superconductivity.  In our case, that scale will be $\langle  \ocal_i \rangle$. 
One important and subtle issue in understanding this equation is that the two limits $\omega \to 0$ and $k \to 0$ do not always commute.  In the limit $k=0$ and $\omega \to 0$, we can take a time derivative of both sides to find 
\be
  J_i (\omega,0)  = \frac{i n_s}{\omega} E_i(\omega,0)
\label{electriclondon}
\ee
explaining the infinite DC conductivity observed in superconductors.  On the other hand, in the limit
$\omega=0$ and $k \to 0$, we can instead consider the curl of the London equation, yielding
\be
i \epsilon_{ijl} k^j J^l(0,k) = - n_s B_i(0,k) \ .
\label{magneticlondon}
\ee
Together with Maxwell's equation $\epsilon^{ijl} \partial_j B_l = 4 \pi J^i$, this other limit of the London equation implies that magnetic field lines are excluded from superconductors.

Thus far in the paper, we have explored the first limit, having set $k=0$ and explored the frequency dependence of the conductivity.  We would now like to argue that the London equation holds more generally, including in the limit where $\omega$ is sent to zero first.  To make life easier, in this section we shall work in the probe limit ($q \to \infty$) in which the metric is kept fixed to be simply the Schwarzschild AdS black hole.  In the probe limit, the scalar and Maxwell field form a decoupled Abelian-Higgs system in this background.  As mentioned previously in section \ref{sec:conductivity}, by decoupling the metric fluctuations, we will remove the additional divergence in the conductivity at $\omega \to 0$ due to translation invariance.  The background metric is 
\be
ds^2 = -g(r) dt^2 + \frac{dr^2}{g(r)} + r^2 (dx^2 + dy^2) \ ,
\ee
where $g(r) = r^2 - r_+^3/r$.

Assume that we have solved self-consistently for $A_t$ and $\psi$ in this Schwarzschild background.  We then allow for perturbations in $A_x$ that have both momentum and frequency dependence of the form $A_x \sim e^{-i \omega t + i k y}$. We have taken the momentum in a direction orthogonal
to $A_\mu$. This allows us to consistently perturb the gauge field without sourcing any other fields.
With these assumptions, the differential equation for $A_x$ reduces to
\be
\left( \frac{\omega^2}{g} - \frac{k^2}{r^2} \right) A_x + (g A_x')' = 2 q^2 \psi^2 A_x \ ,
\label{probemaxwell}
\ee
where $'$ denotes differentiation with respect to $r$.  Ignoring the radial dependence, this equation describes a vector field with mass proportional to $q^2 \psi^2$.  This mass, which is symptomatic of an underlying bulk Higgs mechanism, should give rise to the usual effects of superconductivity, but the radial dependence and the AdS/CFT dictionary cloud the intuition.  

To clarify the situation, first note that we have already explored this equation (\ref{probemaxwell}) numerically in the case $k=0$ both here in section \ref{sec:conductivity} and also previously in \cite{Hartnoll:2008vx}.  In section \ref{sec:conductivity}, strictly speaking we were not working in the probe limit, but we found that as $q \to \infty$, the results approached those of \cite{Hartnoll:2008vx} where we were indeed working in the probe limit.  The results indicated the imaginary part of the conductivity had the form (\ref{electriclondon}), verifying the London equation in the limit $k=0$ and $\omega \to 0$. 

The next observation is that given the structure of (\ref{probemaxwell}), the limits $\omega \to 0$ and $k \to 0$ must commute.  
To compute $n_s$, we simply set both $\omega$ and $k$ to zero and solve (\ref{probemaxwell}).
Thus in the probe limit we directly obtain the magnetic London equation (\ref{magneticlondon}).
Away from the probe limit, these limits may fail to commute. As we saw
in section \ref{sec:conductivity}, Maxwell's equation for $A_x$ (\ref{eq:axprelim}) depends also on the metric fluctuations $g_{tx}$.  We saw that for $k=0$ and $\omega \neq 0$, the metric fluctuations could be replaced by an additional effective mass term for $A_x$.  However, for $\omega=0$ and $k \neq 0$, more metric and gauge field fluctuations were sourced, and the differential equation governing $g_{tx}$ are no longer first order. 

We now attempt to approximate $n_s$ analytically at very low temperature, following a method
that was used successfully in \cite{Hartnoll:2008vx}. 
Without knowing an analytic form for $\psi$, it is not possible to provide an exact solution to this differential equation for $A_x$.  Nevertheless, we find numerically that in the dimension one case $\sqrt{2} \psi \approx \langle \ocal_1 \rangle /r$ to a good approximation everywhere. 
 At very low temperature, $r_+ \to 0$ and our background approaches AdS in Poincar\'e coordinates. Introducing a new radial variable $z=1/r$,  eq.\ (\ref{probemaxwell}) reduces to the Klein-Gordon equation with mass proportional to $\langle \ocal_1 \rangle$:
\begin{equation}
(\omega^2 - k^2 - q^2 \langle \ocal_1 \rangle^2) A_x + \ddot A_x = 0 \ .
\end{equation}
Here a dot denotes differentiation with respect to $z$.
We are implicitly working at low frequencies where  $\omega^2$, $k^2 \ll  q^2 \langle \ocal_1 \rangle^2$.  Since the horizon is at large $z$, we impose the boundary
condition that $A_x$ be well behaved there to find
\begin{equation}
A_x = a_x e^{-i \omega t + i k y - \lambda z} \,,
\end{equation}
where $\lambda^2 = q^2 \langle \ocal_1 \rangle^2 + k^2 - \omega^2 \approx q^2 \langle \ocal_1 \rangle^2$. 

To obtain the conductivity we expand $A_x$ near the boundary $z=0$ in the low frequency case:
\be
A_x = a_x(1 -  \lambda z + O(z^2)) \ .
\ee
From the AdS/CFT dictionary, described several times above, we can interpret the zeroth order term as an external field strength and the linear term as a current $J_x$.  Thus, this expansion gives us a modified London equation:
\be
J_x = - \sqrt{q^2 \langle \ocal_1 \rangle^2 + k^2 - \omega^2} \, \, a_x \ .
\ee
In the limit $\omega$, $k \ll q \langle \ocal_1 \rangle$, we get precisely the London equation:
\be\label{eq:goodlondon}
J_x = - q \langle \ocal_1 \rangle \, a_x \ .
\ee
We have verified numerically that the strength of the pole in the imaginary part of the conductivity is indeed very close to $q \langle \ocal_1 \rangle$ at low temperatures.

A similar estimate of $n_s$ for the $\langle \ocal_2 \rangle$ theory is given in Appendix B. Since the approximation $\sqrt{2} \psi \approx \langle \ocal_2 \rangle /r^2$ is not as good, the estimate is off by about 25\%. But the main point is that by allowing for both a momentum and frequency dependence, we are free to choose the order of limits in which we send the frequency and momentum to zero. If we set $\w=0$ first, then we are manifestly
dealing with a purely magnetic external field. The London equation (\ref{eq:goodlondon})  is precisely what we need to describe both the expulsion of a magnetic field from the superconductor and the infinite conductivity. 

The London equation leads to the magnetic penetration depth
\be
\lambda^2 = {1\over 4\pi n_s} \,,
\ee
via the Maxwell equation for the curl of the magnetic field:
\be
-\nabla^2 B = \nabla \times (\nabla \times B) = 4 \pi \nabla \times J = - 4 \pi n_s \nabla \times A = - 4 \pi n_s B \,.  
\ee
Therefore
\be
\nabla^2 B = \frac{1}{\lambda^2} B \,,
\ee
implying that static magnetic fields can penetrate a distance $\lambda$ into the superconductor.
Although we argued above that a 2+1 dimensional superconductor cannot expel a perpendicular magnetic field, this lengthscale will still play an important role in a gauged extension of our model.
For instance, at low temperatures we expect the flux to be confined to vortices, and their size will
be determined by $\lambda$. Some comments on weakly gauging a holographic superconductor appear in section 7 below.

\subsection{The superconductor in a finite magnetic field}

In the previous section, we studied the onset of superconductivity in the
presence of a constant background magnetic field. We found that at sufficiently
low temperature there is a second order transition to a superconducting droplet.
Thus the first consequence of the magnetic field is to confine the superconducting
condensate to a finite region. In this section we are interested in characterising
the currents associated with the droplet phase.

The formalism here is similar to that used in section \ref{sec:Binstab}.  
Let us write the background metric in polar boundary coordinates:
\begin{equation}
ds^2 = - g(r) dt^2 + \frac{dr^2}{g(r)} + r^2 \left(du^2 + u^2 d\vp^2
\right) \,.
\end{equation}
We will choose a gauge in which the scalar $\psi $ is real.
We are interested in static, axisymmetric solutions in which all fields are
independent of $t$ and $\vp$. In this case, it is consistent to set $A_r = A_u = 0$. The Maxwell and scalar field equations become
\be\label{Ateq}
{1\over u} \pa_u(u\pa_u A_t) + g\pa_r(r^2 \pa_rA_t) = 2r^2 q^2  \psi^2 A_t\ ,
\ee
\be\label{Aphieq}
u\pa_u ({1\over u}\pa_uA_\vp)+ r^2 \pa_r(g \pa_r A_\vp) = 2 r^2 q^2  \psi^2 A_\vp \ ,
\ee
\be\label{psieq}
{1\over u}\pa_u(u\pa_u\psi) + \pa_r(r^2g\pa_r\psi) + \left[{q^2 r^2\over g} A_t^2 - {q^2\over u^2 }A_\vp^2 + 2r^2 \right] \psi = 0 \ .
\ee
To study the complete effects of a magnetic field on the superconductor, one would have to solve these coupled nonlinear partial differential equations. This will require more sophisticated numerical methods.

We will now study these equations in two different limits in which they
become tractable. Firstly we look at the case where the magnetic field is small
compared to the background charge density. In this region we can treat the
magnetic field as a perturbation and linearise. Secondly, we look at temperatures just
below the formation of the superconducting droplet. Here we can treat the scalar field
as a perturbation. Finally we will put together the main features of these two limits
to obtain a qualitative picture of the full phase diagram.

\subsubsection{Small magnetic fields}

Consider the case that the magnetic field is weak and can be treated as a perturbation
of the solution with no magnetic field. We shall see that a 
static magnetic field generates currents in the superconductor.

Assume that we have solved self-consistently for $A_t$ and $\psi$ in the absence of a magnetic field, as we have been doing in earlier sections of this paper. We introduce a small
but nonzero $A_\vp$ in this background. The equation (\ref{Aphieq}) can now be separated $A_\vp =  u V(u) S(r)$ with
\be
V^{''} + {1\over u} V' + \left(c^2 - {1\over u^2} \right)V =0 \ ,
\ee
\be\label{Seq}
(gS')' - \left({c^2\over r^2} + 2q^2 \psi^2 \right) S =0 \ ,
\ee
where $c$ is a constant of integration. The solution for $V$ which is regular at the origin is just the Bessel function $J_1( cu)$ where $c$ is real.  Note that $c$ can be chosen at will and its value determines how spread out the magnetic field is on the boundary.
As $c \to 0$, there is an arbitrarily large region centered at the origin where $A_\vp \propto u^2$, corresponding to a uniform magnetic field.

We now need to solve for $S$ in order to find the current.
One cannot solve for $S$ analytically, but as usual, at large $r$, $S = S^{(0)} + S^{(1)}/r$ where $S^{(i)}$ are constants and
\be\label{eq:londonphi}
J_\vp =  u J_1( cu) S^{(1)} = \frac{S^{(1)}}{S^{(0)}} A_\vp \ .
\ee
  In the absence of a condensate, it is easy to see that the current vanishes.
  Subtracting a constant from (\ref{Seq}) we can view this as an equation for $\tilde S= S - S^{(0)}$ which
  vanishes at infinity. If $\psi =0$, we can multiply (\ref{Seq}) by $\tilde S$ and integrate over the region
  outside the horizon. The result is a nonpositive integrand which integrates to zero, implying $\tilde S =
  0$. 
  
 The equation (\ref{Seq}) is identical to the Maxwell equation (\ref{probemaxwell}) we wrote previously, with $\omega=0$ and $k^2 \to c^2$.  In fact, the only difference between this calculation and the one in section (6.1) is that previously we used cartesian coordinates and a standard plane wave dependence for the magnetic field,  while here we used polar coordinates and the associated Bessel function dependence.  Therefore at low temperature, the coefficient of the London equation (\ref{eq:londonphi})  in the limit $c \to 0$ will be identical to that of the previous subsection. Of course,  since we are not trying to solve for $S$ analytically, we can now consider all $T < T_c$, whereas previously we were restricting attention low temperatures.

\subsubsection{Small condensates}

We now revisit the calculation in section \ref{sec:Binstab} where we examined the limit
where $\psi$ is very small, i.e.\ we are near $T=T_c$. 
The right hand sides of (\ref{Ateq}) and (\ref{Aphieq}) vanish, so $A_\mu$ satisfies the source-free Maxwell equation. One solution to (\ref{Aphieq}) is then simply
\be\label{uniformB}
A_\vp = \frac{B u^2}{2} \,,
\ee
corresponding to a uniform magnetic field. Similarly, $A_t$ can be just a function of $r$. 
As before, the equation for $\psi(r,u) = R(r) U(u)$ can be separated, yielding the differential equations
(\ref{eq:Ueq}) and (\ref{eq:Req}).  The solution for $U$ is a Gaussian profile as in (\ref{eq:gaussian}).
Even though the condensate is nonzero, there is no current generated in this limit because we are working to first order in $\psi$ and  $A_\vp$ is independent of $r$. 

We now wish to support our general picture of currents induced by the external magnetic field by extending our small $\psi$ expansion to ${\cal O}(\psi^2)$ in order to compute the induced current.
Note that unlike in section 5 above, we are in the probe limit here.
Substituting (\ref{uniformB},\ref{psisep}) into the right hand side of (\ref{Aphieq}) we get
\be\label{secondorder}
u\pa_u ({1\over u}\pa_u \delta A_\vp)+ r^2 \pa_r(g \pa_r \delta A_\vp) =
 q^2 Bu^2 e^{-qBu^2/2}  r^2 R^2(r) \ .
\ee
This equation doesn't separate, but we believe that we can understand its solutions as follows.
Immediately we can see that there is a source term for $A_\vp$ that will cause an
$r$ dependence in the solution and hence produce a current in the field theory.
Note that regularity near $u=0$ requires that for small $u$
\be
\delta A_\vp =  u^2  S(r) \ .
\ee
This already shows that the current (\ref{eq:londonphi}) grows with radius $u$ away from the center of the condensate. To understand the structure of the solution at large $u$, we try
\be
\delta A_\vp =  u^p e^{-qBu^2/2}  S_p(r) \ .
\ee
The first term in (\ref{secondorder}) produces terms with $u$ dependence given by the exponential times $u^{p+2}$, $u^p$ and $u^{p-2}$. Clearly $p$ cannot be positive since there are no terms to cancel the highest power of $u$. Setting $p=0$ leads to the approximate  large $u$ solution
\be
\delta A_\vp = {1\over B} e^{-qBu^2/2} r^2 R^2(r) \ .
\ee
This solution can be systematically corrected by  taking
\be
\delta A_\vp = \sum_{m=0}^{\infty} u^{-2m} e^{-qBu^2/2} S_{2m}(r) \ .
\ee
The functions $S_{2m}$ are determined iteratively and algebraically in terms of the lower $m$ functions and their derivatives. This shows that the current dies off exponentially far from the condensate, as one expects, since persistent currents must vanish when there is no condensate.
To make this discussion more rigorous, one should show that there is indeed a solution to the equations matching these two asymptotic behaviours.

\subsubsection{Comments on the full phase diagram}

By piecing together the results from the two limits we have just discussed, small magnetic field and small condensate, we arrive at the following picture for the superconductor in a constant external magnetic field, i.e. without dynamical photons. Because we have not solved the full equations, the description that follows is a minimal interpolation between the different regions we have studied. This phase diagram is sketched in figure \ref{fig:PhaseDiagram} below, with the regions that are accessible to a linearised analysis shaded.

\begin{figure}[h]
\centerline{ \epsfig{figure=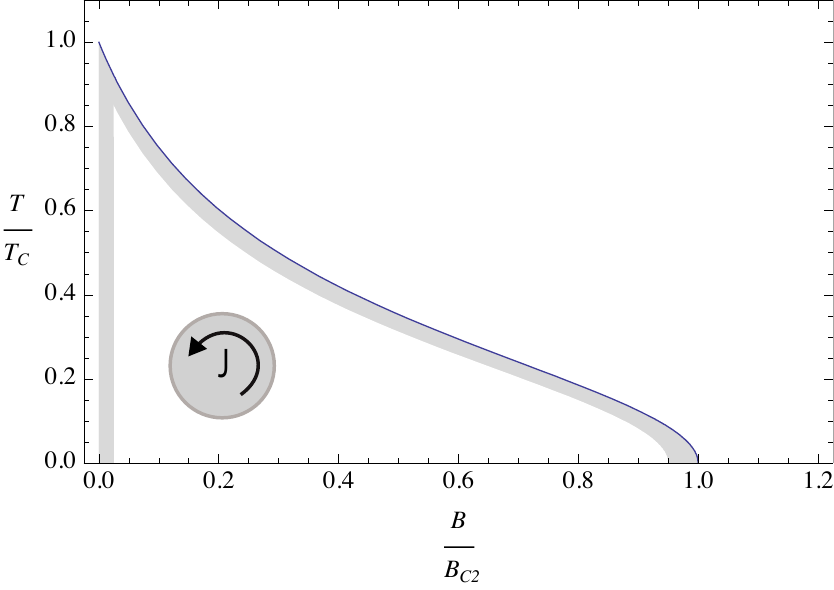, width=4.5in} }
\caption{
\label{fig:PhaseDiagram} Schematic illustration of the full phase diagram of the theory
at finite temperature and external magnetic field. The shaded regions indicate where we
can obtain a description of the phase by linearising the full equations in either the
magnetic field or the condensate. In the low temperature, low magnetic field phase,
the condensate is concentrated in a circular droplet with circular currents in
the superconducting region.
 }
\end{figure}

Firstly, for any finite magnetic field the superconducting condensate will be localised to a finite circular region. As the magnetic field becomes smaller the region grows until it occupies the whole plane in the $B \to 0$ limit. In general there are exponential tails of superconductivity reaching out to infinity. It would be interesting to ascertain whether in the zero temperature limit the condensate becomes completely localised.

Secondly, for any finite superfluid density and magnetic field there are always circular currents generated. The currents are largest just inside the boundary of the condensate and die off exponentially at large radii. When coupled to dynamical photons, these currents act to expel the applied magnetic field.

\setcounter{equation}{0}
\section{Photon mass, symmetry breaking and infinite conductivity}
\label{sec:mass}

In this section we will make some more formal observations about holographic superconductors.
In particular, we wish to give an interpretation of the pole in the conductivity at $\w=0$ in the superconducting phase. By explicitly
coupling the theory to a photon, we will show that the pole is directly related to the photon becoming
massive.\footnote{%
 There is a temptation to identify the pole at $\w=0$ in the conductivity as being due to the
 Goldstone boson of the spontaneously broken (global) $U(1)$ symmetry. This is not correct however.
 The Goldstone boson is manifested as
 a pole in the retarded Greens function of the current at $\w^2 - k^2 = 0$.
 Whether this pole persists at $\w = k = 0$ depends on the direction in which zero is approached in the 
 $(\w,k)$ plane. We are setting $k=0$ first. In this order of limits there is not a pole at $\w=0$. This is
 manifest in our results: if there had been a pole in the retarded Greens function at $\w=0$, then the
 conductivity would have had a double pole at $\w=0$, because of the relation $\sigma(\w) = - i
 G^R_{J_x J_x}(\w)/\w$ [From Ohm's law:  $J_x = \sigma E_x = i \w \sigma A_x = G^R_{J_x J_x} A_x$.]
}

As we have mentioned several times by now, the 2+1 dimensional
theory we have been considering does not have
a dynamical photon. The $U(1)$ symmetry that is
spontaneously broken is global. Although the dynamics of
spontaneous symmetry breaking does not depend on the photon, much of the interesting
phenomenology of superconductors is concerned with the interaction
of the theory with a dynamical photon.

The 2+1 theory can be coupled to a photon through the standard
$J_\mu A^\mu$ interaction. To make the photon dynamical, we can
add an $F^2$ term to the action, with $F=dA$. Electromagnetic phenomena
such as screening are determined by the effective action for the
photon. We can obtain this action by integrating out all the other
degrees of freedom. In terms of the Euclidean partition function, we have
\bea\label{eq:Z}
Z & = & \int \calD A \calD X e^{ - S[X] - \frac{1}{4 e^2}\int
d^3x F_{\mu \nu} F^{\mu \nu} - \int d^3x J_\mu A^\mu}  \\
 & = & \int \calD A e^{- S_\text{eff.}[A]} \,.
\eea
In these expressions $X$ denotes the degrees of freedom in the 2+1
dimensional theory. Unlike in previous sections, in this discussion
we are taking the photons to also be 2+1 dimensional. This is
because we are not interested in mimicking experimental setups, but
rather in demonstrating a formal property of the theory. Note that
the coupling $e^2$  is therefore dimensionful.

Up to quadratic order in the Maxwell field,
the effective action can be straightforwardly obtained by expanding out the
exponent in (\ref{eq:Z}), integrating over $X$, and then
re-exponentiating. For our theory, in which there is a charge
density but no background currents, we get
\be\label{eq:effective}
S_\text{eff.}[A] =  \frac{1}{4 e^2}\int
d^3x F_{\mu \nu} F^{\mu \nu} +
\frac{1}{2} \int d^3x d^3y \langle J^\mu J^\nu \rangle_c (x-y) A_\mu(x) A_\nu(y)
+ \int dt \, Q \mu
\,.
\ee
This expectation value is evaluated in the theory without
dynamical photons. The last term arises from setting
\be
\int d^3x \rho A_t(x) = \int dt \, Q \mu \,.
\ee
This term is an overall constant, does not affect the dynamics, and so
we shall drop it from this point on.

In expanding the effective action in powers of $A$ one might worry
about the fact that we have dropped the higher order, interacting terms. 
We are helped here by the large $N$ limit, which is a classical
limit of the theory. The coefficients of the higher order terms in $A$
in the action are given by connected higher $n$-point functions of the current:
$\langle J^n \rangle_c$. Connected diagrams arise upon re-exponentiating
the partition function (\ref{eq:Z}) after integrating out the CFT degrees
of freedom. These connected diagrams are computed in AdS/CFT by using
Witten diagrams in the bulk theory which have vertices or loops
or both. Therefore they are suppressed by inverse powers of $N$
compared to the disconnected diagrams at the same order. If we were to
rescale $A$ so that the coefficient of the $A^2$ term in the action were $\ocal(1)$, then
the higher order terms in the effective action would be suppressed by powers
of $N$. 
In this
paper we are studying the theory as a function of frequency at
zero spatial momentum. 
For the $F^2$ term in the action also to be bigger than these interaction
terms we need $\w^2/e^2 \gtrsim \int d^3x \langle JJ \rangle_c$.

An immediate feature of the effective action (\ref{eq:effective})
is the possible presence of a mass term for the photons, via
$\langle J^\mu J^\nu \rangle_c$. Let us see if we can
extract the photon mass. Lorentz invariance is broken by the charge
density, so we need to define what we mean by the photon mass.
From our results we can obtain the energy of photons in a
frame where they are at rest relative to the background charge
density. It is reasonable to associate this energy with the mass of the photons.
Therefore, to find the photon mass we need to exhibit an
on shell photon mode with $k = 0$. The energy $\w$
of this mode will be the photon mass, $m_\gamma$. An example
of a physical consequence of this definition of the photon mass
is that generic zero momentum processes involving massive photons will
decay in time like $e^{- m_\gamma t}$.

Because we have a quadratic effective action, the spectrum can be
obtained directly from the classical equations of motion.
It is straightforward to obtain the equations of motion that follow from
the effective action (\ref{eq:effective}). Let us note that it is consistent to
restrict to a mode in which $A_x$ is the only nonvanishing
component and where
\be\label{eq:BB}
A_x(\w) = y B e^{-i \w t} \,.
\ee
The $y$ dependence has been included so that the mode is not
pure gauge at $\w=0$. Rather, it reduces to a constant magnetic field.
We then Wick rotate so that we are in Lorentzian signature.
The equation of motion for this mode is
\be\label{eq:onshell}
\left(\w^2 + e^2 G^R_{J^x J^x}(\w) \right) A_x(\w) = 0\,.
\ee
Here we used the fact that the retarded Green's function in
momentum space is the Fourier transform of the Euclidean
Green's function in momentum space, up to possible contact terms. However,
it was shown in  \cite{Hartnoll:2007ip} that the contact terms
are not present in the case of the current-current correlator
in this theory.  We are also assuming that $G^R_{J^xJ^y}=0$.

Using  $G^R_{J^x J^x}(\w) = i
\w \sigma(\w)$, we can recast the `dispersion relation' as
\be\label{eq:dispersion}
\w \left(\w + i e^2 \sigma(\w) \right) = 0 \,.
\ee
The solution to this equation gives the photon mass $m_\gamma =
\w$. We can see immediately that $\w = 0$ is a solution
provided that the conductivity does not have a pole (or worse)
as $\w \to 0$. In particular, if the conductivity tends to a
constant or vanishes in the $\w \to 0$ limit, then the photon is
massless. In Lorentz invariant theories we expect the photon to be
massless in phases where the electromagnetic $U(1)$ is unbroken.
Conversely, we expect the photon to gain a mass if the
electromagnetic symmetry is broken. If
the imaginary part of $\sigma(\w)$ has a pole as $\w \to 0$, the photon
will have a nonzero mass. It is important to note, however, that
we also have a medium, the charge density, which breaks
Lorentz invariance.

How do these expectations compare with our results? Take the
superconducting phase first. There indeed the imaginary part of
the conductivity has a pole at the origin, and so $\w=0$ is not a solution to
the dispersion relation (\ref{eq:dispersion}), and hence the photon is
massive, perfectly consistent with expectations. We found above
that $\sigma(\w) \sim i n_s/\w$ as $\w \to 0$  (this is the
definition of $n_s$ for us). This behaviour is not exact. At zero
temperature, for instance, the imaginary part of the conductivity
goes to zero at the gap $\w = \w_g$. In general we will have to solve
(\ref{eq:dispersion}) numerically. However, if $e$ is small enough
we will have
\be
m_\gamma = e \sqrt{n_s} \qquad \text{provided} \qquad e \sqrt{n_s} \ll \w_g \,.
\ee

In the probe limit (recall this was $q \to \infty$), the conductivity
$\sigma$ is constant in the normal phase. There is no pole as
$\w \to 0$ and therefore we find that the photon is massless, 
again consistent with our naive expectations for a symmetric phase.

Beyond the probe limit, when back reaction of the scalar field on
the metric is taken into account we found that the conductivity also
diverged as $\w \to 0$ above $T_c$.  This pole in the conductivity 
was not due to spontaneous symmetry breaking, but rather due
to a translationally invariant charged medium and can be moved away
from $\omega=0$  by impurities or by considering finite momentum.
A charged medium often results in screening of electromagnetic fields,
and our pole at finite $\w$ is consistent with this.
We should also note that the limits $\w \to 0$ and $B \to 0$ do not
commute when back reaction is included \cite{Hartnoll:2007ip, Hartnoll:2008hs}
and so it is subtle to take simultaneously the small $\w$ and $B$ limits in this case.

\setcounter{equation}{0}
\section{Comparison to Landau-Ginzburg}

Before concluding we would like to discuss the microscopic status of
our phenomenological holographic superconductor, in particular
the extent to which it is similar and distinct from a dressed-up
version of Landau-Ginzburg theory. At a first glance what we are doing feels
a lot like Landau-Ginzburg theory --- we have studied a theory like the Abelian-Higgs model,
albeit in one higher spacetime dimension and with a dynamical black hole metric,
in which there is a condensate for a charged complex scalar field. Furthermore,
in our work there has been no sign of what one traditionally expects to find
in a microscopic description of superconductivity: a discussion of a pairing mechanism and
a Lagrangian for the degrees of freedom that form the Cooper pairs.

Recall that Landau-Ginzburg theory is the effective field theory description of
superconductors near  the superconducting phase transition.
The dynamical degree of freedom is the complex order parameter, $\varphi =
\langle \ocal \rangle$, that is coupled to
a background electromagnetic field. The free energy density in the order parameter is
\be
\Delta f_\text{L-G} = \frac{1}{2 m^*} |(\nabla + i q A) \varphi |^2 + a |\varphi|^2
+ \frac{b}{2} |\varphi|^4 \,.
\ee
In this expression $m^*, q, a$ and $b$ are phenomenological parameters. The quantity
$a$ goes through zero at the critical temperature $a \sim \dot a (T - T_c)$. From this
free energy one can derive important quantities such as the superconducting coherence length, $\xi$,
and the superfluid density
\be
\xi \sim \frac{1}{(a m^*)^{1/2}} \,, \qquad n_s \sim \frac{a}{b} \,.
\ee

Landau-Ginzburg theory is not a microscopic theory. If we wished to study low temperatures,
away from the critical temperature, we would have to supplement the above expression
with an infinite number of coefficients describing a general functional of $\varphi$.
The curve $|\varphi(T)|$, for instance, is an input to rather than an output from this functional.
For that one needs BCS or some other microscopic theory.
The usefulness of Landau-Ginzburg theory near the critical temperature is that it
relates various experimental quantities and can describe the interaction of a 
superconducting condensate with an electromagnetic field.

The structural similarity with our phenomenological holographic theory is that we also have
an infinite number of undetermined parameters at low temperatures. We chose a potential
with only a mass term, but we could have chosen an arbitrary function of $\psi$.
We could also have taken a nonminimal coupling between the scalar and the Maxwell field.
Although this question remains to be fully investigated, we believe that our numerical results
for $\langle \ocal (T) \rangle$, for instance, will depend significantly
on the gravitational action. As with Landau-Ginzburg theory, near the critical temperature our
model is much more constrained --- only the mass term is important to lowest order.

Despite these similarities, there are three important differences.
First, the instability which leads to the superconducting phase transition in the CFT has a more natural interpretation in the gravity theory than in Landau-Ginzburg theory.  Gravitationally, given a charged black hole and a fixed mass scalar field, the scalar will typically develop a nontrivial profile at sufficiently low temperature.  Moreover, the curvature of the geometry stabilizes the instability without need for higher order terms in the scalar potential.  In contrast, in Landau-Ginzburg theory, a temperature dependent mass term is added by hand and then stabilized by an additional quartic interaction.



Second, there is a natural way to promote our phenomenological holographic superconductor into a full microscopic description: If we had realised our model as a limit of string theory, then
the potential for $\psi$ would be completely fixed and there would be no free
parameters. We would have a concrete CFT that underwent a superconducting phase
transition at a critical temperature specified by the background charge density.
Furthermore, in this theory, the AdS/CFT correspondence allows us to compute
all the quantities for this superconductor which would normally follow from a BCS-like
treatment: the gap as a function of temperature, the frequency dependent conductivity,
the magnetic penetration depth, etc. We have shown how to use AdS/CFT to compute
these quantities in this paper and we see that the `feel' of the computation is completely
different from weakly coupled BCS-like theories.
Nonetheless, AdS/CFT applied to a model embedded in string theory would be an
honest-to-goodness microscopic computation of these quantities in a well-defined theory.

Thirdly, the physical meaning of the potential appearing in the bulk
description is completely different from the potential in Landau-Ginzburg theory.
The AdS/CFT dictionary repackages the degrees of freedom of the CFT to make manifest the
classicality of a large $N$ limit. The theory remains strongly coupled. AdS/CFT is not
effective field theory. Our phenomenological choice of a `minimal' model is not guided by
Wilsonian arguments but rather by simplicity in terms of the degrees of freedom arising
through the AdS/CFT dictionary. This approach makes sense if one accepts the
AdS/CFT correspondence as the natural tool for an analytic description of strongly coupled
theories. The ultimate test of this assumption will be the success and
robustness of predictions from phenomenological AdS/CFT
in modeling superconductivity in experimental systems where strong coupling
and perhaps scale invariance (quantum criticality) play a key role.
The heavy fermion compounds come to mind as good candidate systems \cite{heavy}.

\setcounter{equation}{0}
\section{Summary}

The main points we have made in this paper are as follows:

\begin{itemize}

\item A minimal AdS/CFT superconductor has a bulk description with a metric, a Maxwell field and a charged scalar field (section \ref{sec:bulk}).

\item If the CFT is placed at a finite charge density and if the scalar is sufficiently light and/or sufficiently
charged, then there is a charged condensate in the theory below a critical temperature, $T < T_c$. We noted there are two distinct reasons why this condensation can happen (section \ref{sec:phases}).

\item We computed the charged VEV as a function of temperature (figure \ref{condensate}) and the critical temperature as a function of the charge $q$ of the operator that condenses (figure \ref{Tc}).

\item We computed the frequency dependent conductivity $\sigma(\omega)$ (figures \ref{realsliceplots} and \ref{lowtempplots}) and showed that a gap opens up for $T < T_c$. There is a delta function in ${\rm Re}\  [\sigma]$ at $\omega = 0$ corresponding to an infinite DC conductivity. We also obtained thermal and thermoelectric conductivities (section 4).

\item We studied the effect of adding a magnetic field to the holographic superconductor and argued that they are always type II. Superconducting droplets form as the magnetic field is lowered (section 5). 

\item We showed that holographic superconductors generate screening currents obeying the London equation. A photon coupled to the superconductor acquires a mass (sections 6 and 7).

\item Using gravity to study superconductivity has a superficial similarity to a Landau-Ginzburg description. Key differences include the fact that the phase transition does not have to be put in by hand, and a complete microscopic theory could be obtained by realizing a model similar to ours as a limit of string theory (section 8).

\end{itemize}

The are many remaining questions to address. Besides technical issues such as understanding the zero temperature limit better, three pressing directions of research might be emphasized: Firstly to find robust (`universal') results from phenomenological holographic superconductors and to understand the extent to which these are useful results for experimental systems
involving nonconventional superconductivity; secondly, to obtain a microscopically understood model by embedding a holographic superconductor into string theory; and thirdly to find new mechanisms for superconductivity in the AdS/CFT correspondence, perhaps with d-wave symmetry or where the critical temperature $T_c$ is set by a dynamical scale $\Delta$ rather than a charge density.

\vskip 1cm
\centerline{\bf Acknowledgements}

We would like to thank Andrei Bernevig, David Huse, Andreas Ludwig, Rafael Porto and Matt Roberts for
illuminating discussions.  SAH was based at the KITP in Santa
Barbara while much of this work was done. SAH and CPH would acknowledge the stimulating
hospitality of the Tata Institute, Mumbai, while part of this work was underway.
This work was supported in part by NSF grants PHY-0756966, PHY-0555669 and PHY05-51164.

\appendix

\section{Instability with a neutral scalar field}
\label{app:unstable}

In this appendix we give a proof of the following claim made in
the text: When coupled to a neutral scalar field with $m^2 = -2$,
the Reissner-Nordstrom AdS black hole becomes unstable near
extremality. We shall prove this using test functions and the
Rayleigh-Ritz method.

Let us write the black hole metric in terms of the coordinate
$z=1/r$
\be
ds^2 = \frac{1}{z^2} \left[ -f(z) dt^2 + \frac{dz^2}{f(z)} + dx^2 + dy^2 \right] \,.
\ee
where
\be
f = 1  - \left(1 + c^2\right) z^3 + c^2 z^4
\,.
\ee
Here $c$ denotes a dimensionless charge density, obtained by rescaling the
horizon to $z=1$. It is related to the physical charge density $\rho$ and
temperature $T$ by \cite{Hartnoll:2007ip}
\be
\frac{\rho}{T^2} = \frac{16 \pi^2 c}{(3-c^2)^2} \,.
\ee

We now want to write the equation of motion for the neutral scalar field $\psi$ in
Schr\"odinger form. This form will help us gain intuition about the behaviour of the field.
The rewriting requires rescaling the field and changing variables
from $z$ to a new coordinate $s$. Let
\be
\psi = z \Psi \,, \qquad \frac{ds}{dz} = \frac{1}{f} \,.
\ee
(Although irrelevant for the following argument, it follows that the range of $s$ is from $0$ (the boundary) to $\infty$ (the horizon)).
Then, taking $\Psi(s,t)$ with a time dependence $e^{- i \w t}$, we obtain the Schr\"odinger equation
\be
- \frac{d^2 \Psi}{ds^2} + V(s) \Psi = \w^2 \Psi \,,
\ee
with the potential, written in terms of the $z(s)$ variable,
\be\label{eq:potential}
V = - f \left( \frac{2}{z^2} + \frac{f'}{z} - \frac{2 f}{z^2} \right) \,.
\ee
If this Schr\"odinger equation has a negative energy bound state,
then we have found an instability of the black hole. 
Negative energy in this context implies that $\w^2 <
0$.  Hence $\w$ is pure imaginary and there are solutions which
grow exponentially in time.

This potential (\ref{eq:potential}) is positive everywhere
unless the charge density $c$ is close to the extremal value $c =
\sqrt{3}$ (i.e. $T=0$). More specifically, the potential develops a negative
region in the vicinity of the horizon for $c > 1$. So any
instability is restricted to the charge values $1 < c
\leq \sqrt{3}$, i.e. $4 \pi^2 \leq \rho/T^2 < \infty$.

We cannot solve this Schr\"odinger equation exactly. We solved it numerically in
the main text. However, to show the existence of a negative energy
bound state it is sufficient to find a test function, satisfying
the correct boundary conditions, which gives a negative energy.
The action to use depends on the boundary conditions of the field
$\Psi$. The general allowed falloff at the boundary $z \to 0$ is
\be
\Psi \sim a + b z + \cdots \,.
\ee
We can consider either the first or the second of these terms to be the `non-normalisable' mode.
The action must be stationary under variations of the normalisable mode.
If we impose the boundary condition $\delta a=0$ (i.e. $\delta \Psi(0) = 0$) the following action is stationary on solutions to the Schr\"odinger equation
\be
S_{\delta a=0} = \int ds \left[ \left( \frac{d \Psi}{ds} \right)^2 + \left( V(s) - \w^2\right) \Psi^2 \right] \,.
\ee
However, if we wish to impose $\delta b=0$ (i.e. $\delta \Psi'(0)=0$) then we must add a boundary term
\be\label{eq:azaction}
S_{\delta b=0} = \left. \int ds \left[ \left( \frac{d \Psi}{ds} \right)^2 + \left( V(s) - \w^2\right)  \Psi^2 \right]  -2 \Psi \frac{d \Psi}{ds} \right|_{s \to 0}  \,.
\ee
Both of these actions are finite on solutions to the Schr\"odinger
equation, partly due to a cancellation in the potential
(\ref{eq:potential}) as $z \to 0$ which only occurs at the mass we
have chosen, $m^2=-2$. In fact, the boundary term in
(\ref{eq:azaction}) vanishes on shell for `normalisable' modes.

To show an instability, we need to find test functions such that
$S(\w=0) < 0$. 
Let us start with the second of the boundary conditions above. Normalisable modes therefore have
$b=0$. A simple test function that satisfies this boundary condition is
\be\label{eq:simpletest}
\Psi_\text{test} = 1 - \alpha z^2 \,.
\ee
It is easy to check that the ($\w=0$) action is minimised
by
\be
\alpha = \frac{21 (3 c^2 - 5)}{10 (9 c^2 - 35)} \,.
\ee
This function then leads to an action which is negative for $1.609
\lesssim c \leq \sqrt{3} \approx 1.732$. Note that in this
expression the lower bound in $c$ is an analytic result, it can be
expressed in terms of ratios of square roots. If we take a more
sophisticated test function we can lower the bound a little. For
instance, taking a fourth order polynomial in $z$ instead of
(\ref{eq:simpletest}) leads to the lower bound $c \approx 1.584$
which is close to actual value found numerically in the main text
($c \approx 1.582$). Therefore the test function method not only
indicates the existence of unstable black holes for this boundary
condition, but also gives a good estimate of the minimal
$\rho/T^2$ at which the normal phase is unstable.

We can also consider the boundary condition in which
`normalisable' modes have $a=0$. For this case we have not found a
test function indicating the existence of an instability.
Na\"ively speaking, this is because the falloff $\Psi \sim z$
rather than $\Psi \sim 1$ forces more kinetic energy into the
field. Our full numerics in the main text suggest (or at least,
are consistent with the idea) that there should be an unstable
mode in this case also, but that the critical value of $c$ should
be very close to the extremal value $c=\sqrt{3}$.

\setcounter{equation}{0}
\section{Analytic estimate of $n_s$ for $\langle \ocal_2 \rangle$ case}

In this appendix, we attempt to calculate $n_s$ analytically at low temperatures for the dimension two case, using the method in  section (6.1).
If we assume that at very low temperature, $\sqrt 2 \psi \approx  \langle{\mathcal \ocal}_2 \rangle/r^2$, then (\ref{probemaxwell}) becomes
\be
(\omega^2 - k^2 - q^2 \langle \ocal_2 \rangle^2 z^2) A_x + \ddot A_x  = 0 \ .
\ee
where $z=1/r$ and a dot denotes $d/dz$. This differential equation can be solved in terms of parabolic cylinder functions, $D_{\nu} (c z)$ where the choice
\be
\nu = - \frac{1}{2} + \frac{k^2 - \omega^2}{2 q \langle \ocal_2 \rangle} \ \quad \mbox{and} \quad c = \sqrt{2 q \langle \ocal_2 \rangle} \ 
\ee
gives the proper exponential fall-off as $z$ gets large.  Here the condition that $k$ and $\omega$ are small is more precisely $k^2$, $\omega^2 \ll q \langle \ocal_2 \rangle$.  Expanding $D_{\nu}(c z)$ near the boundary, we find
\be
A_x = a_x \left(1  - \frac{2 \Gamma(3/4)}{\Gamma(1/4)}\sqrt{ q \langle \ocal_2 \rangle} \, z + {\mathcal O}(z^3) \right) \ ,
\ee
where we have suppressed corrections in $(k^2 - \omega^2) / q \langle \ocal_2 \rangle$.  The London equation here is then
\be
J_x = -  \frac{2 \Gamma(3/4)}{\Gamma(1/4)}\sqrt{ q \langle \ocal_2 \rangle}  \, \, a_x
\label{eq:goodlondontwo}
\ee
Numerically, this estimate of $n_s$ appears to be wrong by about $25\%$ at low temperatures.  While $2 \Gamma(3/4)/\Gamma(1/4) \approx 0.676$, the real constant of proportionality appears to be about 0.546.

\end{document}